\documentclass[11pt]{article}
\usepackage[sc]{mathpazo} %Like Palatino with extensive math support
\usepackage{fullpage}
\usepackage[authoryear,sectionbib,sort]{natbib}
\usepackage[utf8]{inputenc}
\usepackage{multicol,caption}
\usepackage[switch]{lineno} 
\usepackage[pdftex]{graphicx} 

\title{Identifying differences in the rules of interaction between individuals in moving animal groups}
\author{Timothy. M. Schaerf$^{1,\dag}$ \\ 
James E. Herbert-Read$^{2,\ast,\dag}$ \\ 
Mary R. Myerscough$^{3}$ \\
David J. T. Sumpter$^{2}$ \\
Ashley J. W. Ward$^{1}$ }
\date{}

\begin{document}
%\linenumbers

\maketitle

\noindent{} 1. Animal Behaviour Lab, School of Biological Sciences, University of Sydney, Sydney, 2006, Australia;

\noindent{} 2. Department of Mathematics, Uppsala University, Uppsala, 75106, Sweden;

\noindent{} 3. School of Mathematics and Statistics, University of Sydney, Sydney, 2006, Australia.

\noindent{} $\ast$ Corresponding author; e-mail: james.herbert.read@gmail.com.

\noindent{} $\dag$ These authors contributed equally to this study.

\bigskip

%\textit{Manuscript elements}: Figure~1, figure~2, figure~3, online appendix~A (including figures~A1: A22 and tables A1:A5). Figure~1,2 and 3 are to print in color.

\bigskip


\bigskip

\textit{Manuscript type}: Article. %Or e-article, note, e-note, natural history miscellany, e-natural history miscellany, comment, reply, invited symposium, or countdown to 150.

\bigskip

%\noindent{\footnotesize Prepared using the suggested \LaTeX{} template for \textit{Am.\ Nat. 
%\
%The authors wish to be identified to the reviewers
%\
%}}

%\linenumbers{}
%\modulolinenumbers[3]

\newpage{}
\begin{multicols}{2}
\section*{Abstract}

\textbf{Collective movement can be achieved when individuals respond to the local movements and positions of their neighbours. Some individuals may disproportionately influence group movement if they occupy particular spatial positions in the group, for example, positions at the front of the group. We asked, therefore, what led individuals in moving pairs of fish  (\emph{Gambusia holbrooki}) to occupy a position in front of their partner? Individuals adjusted their speed and direction differently in response to their partner's position, resulting in individuals occupying different positions in the group. Individuals that were found most often at the front of the pair had greater mean changes in speed than their partner, and were less likely to turn towards their partner, compared to those individuals most often found at the back of the pair. The pair moved faster when led by the individual that was usually at the front. Our results highlight how differences in the social responsiveness between individuals can give rise to leadership in free moving groups. They also demonstrate how the movement characteristics of groups depend on the spatial configuration of individuals within them.}

\section*{Introduction}

Collective motion is often driven through the local interactions between individuals in a self-organising process (\citealt{Couzin2005,Couzinetal2002}). By responding to the movements and positions of their neighbours, individuals can transfer information about detected threats or move together towards target locations (\citealt{sumpter2010collective,herbert2015initiation,couzin2011uninformed}). But information does not always propagate evenly across groups (\citealt{Rosenthal2015rev}). When individuals in bird flocks or fish schools travel in the same direction, individuals positioned at the front of groups are more likely to initiate changes in the direction of others, since individuals cannot observe the movements of those directly behind themselves (\citealt{HerbertReadetal2011,Nagy2010}). If some individuals within a group occupy these front positions more than others, then these individuals can disproportionately influence group motion (\citealt{couzin2011uninformed,reebs2000can}). In effect, minorities of individuals can guide entire groups (\citealt{Couzin2005}). 

There is evidence that individuals consistently occupy different positions within moving groups. Phenotypic characteristics such as body size can determine where individuals are located in groups, with larger individuals sometimes occupying positions at the front of groups (\citealt{pitcher1982evidence}). Differences between individuals' movements may also lead to differential spatial positions, with faster moving individuals migrating towards front positions (\citealt{pettit2013interaction, Couzinetal2002}). However, some individuals occupy positions in groups that do not seem to be related to their body size or other phenotypic differences. Burns et al. (2012), for example, found that individual mosquitofish (\emph{Gambusia holbrooki}) occupied consistent positions within groups, and this was not related to their sex, body size, or dominance (\citealt{burns2012consistency}). Similarly, dominance cannot explain the differential spatial positioning of individuals in homing pigeon flocks (\emph{Columbia livia}) (\citealt{Nagyetal2013}). In these cases, what differences between individuals determine where individuals position themselves in groups?  

Other factors that are more transient within an individual may explain the differential positioning behaviour of individuals in groups. In some cases, individuals that are more informed about their environment than other group members may occupy positions at the front of groups (\citealt{reebs2001influence}) (but see; \citealt{flack2013robustness}). Nutritional state may also affect where an individual positions itself in a group, with hungry individuals often being found at the front of groups (\citealt{krause2000leadership}). In these cases, and in the absence of energetic or phenotypic constraints on movement, it is ultimately how an individual interacts with its neighbours that leads it to occupy different spatial positions. Information and satiation level influence the likelihood that an individual will follow another conspecific's movements (\citealt{nakayama2012initiative,Kingetal2011,leblond2006individual}).  If an individual ignores the movements of others more than others ignore it, then passive self assortment may move this individual to the front of groups (\citealt{Couzin2005}). How an individual interacts with its neighbours, and in particular, whether some individuals are more `socially responsive' to the movements of others, therefore, is likely to result in differential spatial positioning in groups and affect who follows whom (\citealt{harcourt2009social}).   

How can we measure the responsiveness of individuals to each others' movements?  Researchers have previously used averaging procedures to determine the general `rules' that individuals in groups use to react to their neighbours when on the move (\citealt{HerbertReadetal2011,Katzetal2011,Lukeman2010, gautrais2012deciphering}). Attraction and repulsion rules, largely governed by changes in speed, appear important determinants of how individuals respond to their neighbours' locations and movements (\citealt{HerbertReadetal2011,Katzetal2011}).  Alternatively, these rules may be driven by individuals preferring to occupy particular spatial positions with respect to the locations of other group members (\citealt{perna2014note}). The variation in these rules between individuals, however, has not been explored experimentally, although models of collective motion show that differences between individuals' movements and interactions can influence group dynamics (\citealt{romey1996individual,Couzin2005}). 

We tested whether individuals differed in how they responded to their partner's location and movements in pairs of mosquitofish (\emph{Gambusia holbrooki}) during their exploration of an unfamiliar arena.  We first asked whether an individual's position in the group was related to their relative size, movement profile (for example, their average speed) or information about the environment, and then determined whether differences in the social interactions between individuals could lead to individuals occupying different positions within groups. 

\section*{Methods}

\subsection*{Experiments}

Female mosquitofish (n = 80) (\emph{Gambusia holbrooki}) were collected using hand-nets from Lake Northam, Sydney, NSW, ($33^{\circ}53'07''$ S; $151^{\circ}11'35''$ E).  Fish were held in 170 l aquaria and were fed flake food \emph{ad libitum}. Fish were kept for at least 3 weeks prior to experimentation. A square experimental arena (1.5 m x 1.5 m x 0.2 m) was constructed of opaque white perspex and filled to a depth of 7 cm.  In two corners of the arena, diagonally opposite one another, we placed an opaque white holding tube (10 cm diameter). For each trial, we selected two fish of similar size (approximately 1.5 - 2.5 cm) and placed one in each of the holding tubes. To test if a fish's experience with the environment subsequently led this individual to occupy positions at the front (or back) of the group, following an acclimation period of 5 minutes, we either released one (n = 20 trials), or both fish (n = 20 trials) into the arena.  If we had only released one fish into the arena, we allowed this fish to explore the arena for 5 minutes before we released the second fish.  In one of the two treatments, therefore, one fish had explored the environment for longer than the other.  We filmed the trials using a Basler avA1600-65kc camera and recorded using StreamPix (version 5) at 40 fps.  Fish were filmed for 6 minutes when both fish had been released into the arena. These films were subsequently converted using VirtualDub  (version 1.9.11) and the fish were tracked using Ctrax (version 0.5.4), (\citealt{Bransonetal2009}). Any ambiguities in fish identities or other elements of the tracked data were resolved using Ctrax's \emph{fixerrors} GUI. In addition to time series of each fish's $(x,y)$ coordinates, we extracted measurements related to the body length of each fish.

\subsection*{Analysis}

We smoothed the $x$ and $y$ components of each fish's track using a Savitzky-Golay filter (implemented through MATLAB's intrinsic \emph{smooth} function with span 5 and degree 2) prior to all diagnostic calculations. We only analysed basic individual characteristics of each fish's motion, positioning behaviour and directional correlation when fish were less than or equal to 100 mm apart. This is because we wanted to ensure we analysed sequences when the pair was interacting, and not when individuals were exploring the arena independently.  58.49\% of our trajectory data satisfied this condition. Further, our analysis of interactions between individuals was restricted to cases where the relative displacement between the fish was less than or equal to 100 mm in both the $x$ or $y$ direction.  

\subsection*{Predictors of positioning behaviour in pairs}

We first determined which individual in the pair was more often in front of the other (relative to the heading of group motion) over the entire trial. We defined individuals that were more often observed at the front of the group `front fish', and individuals that were more often observed at the back of the group `back fish' (see online appendix A for the proportion of time these individuals occupied different positions). 

We then determined the average directional correlation between the two fish as a function of a delay time to examine the relative influence that fish had on the direction of motion of their partner when they were either at the front or back of the pair (\citealt{Nagy2010,Katzetal2011}). We calculated each fish's speed, magnitude of acceleration, change in speed over time and turning speed for all frames when fish were less than or equal to 100 mm apart. We determined the mean, standard deviation, median, inter-quartile range and maximum value of each of these variables for front fish and back fish separately (details of all the above calculations are provided in online appendix A.) 

%%%%%%%%%%%%%%%%%%%% Figure %%%%%%%%%%%%%%%%%%%%%%
\begin{figure*}
\centering\includegraphics[width=\textwidth]{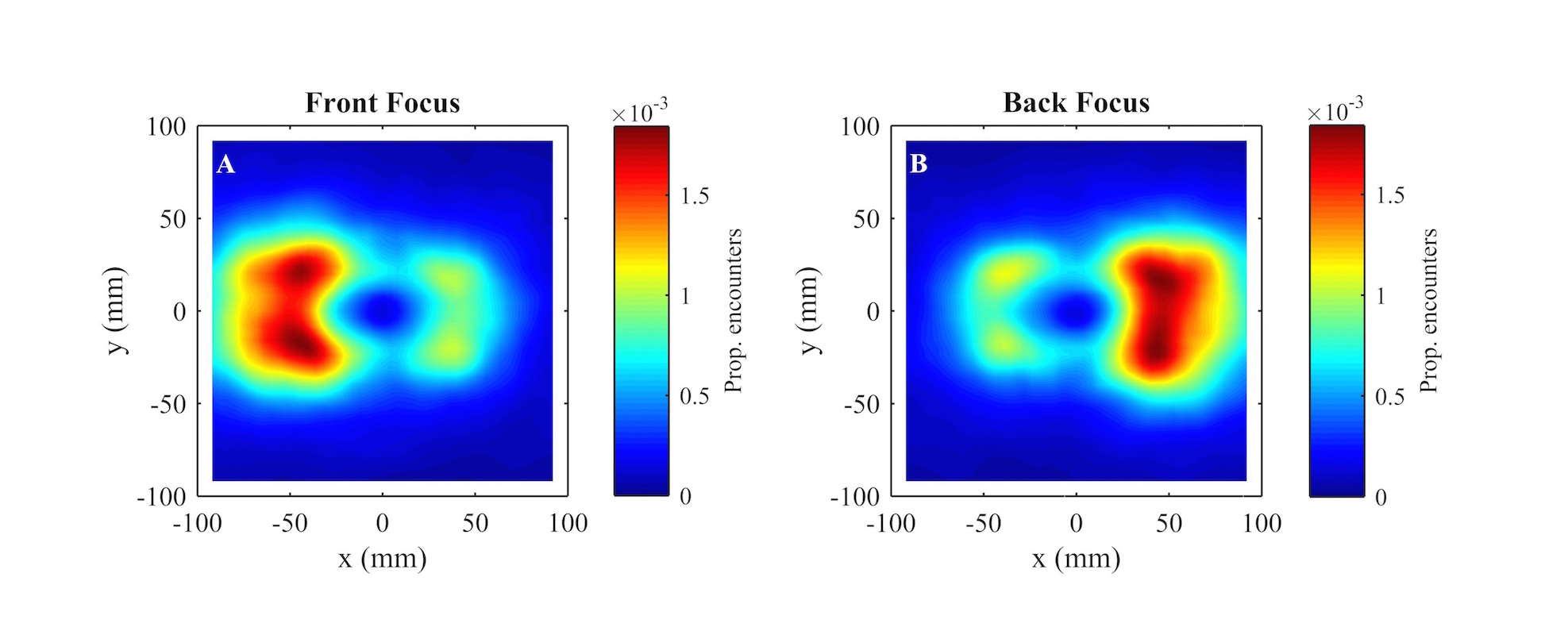}
\caption{Spatial positioning of front and back fish. The heat in the plots shows the proportion of encounters of a neighbour when at different locations relative to the direction of motion of the focal fish. Focal fish are located at the origin of each plot, moving in the direction of the positive $x$-axis. Front fish (A) had their neighbour more often behind them, than back fish (B).}\label{fig:Fig1}
\end{figure*}
%%%%%%%%%%%%%%%%%%%% Figure %%%%%%%%%%%%%%%%%%%%%%

We asked which of the above movement parameters or body size was greater (or less) for front or back fish. To do this, 
we treated all the summary statistics as paired data (front fish versus back fish) and calculated the difference in parameter values between front and back fish. We applied a Shaprio-Wilk test to determine if these differences were likely to have been drawn from a normal distribution \citealt{ShapiroWilk1965}. If the data was normally distributed, we then performed a paired $t$-test to determine if the mean of the differences differed from zero. If the set of differences was not normally distributed, we performed a two-sided Wilcoxon paired-sample test (via MATLAB's intrinsic \emph{signrank} function) to determine if the median difference in parameters differed from zero (see for example, \citealt{Zar1996}). To account for the number of summary statistics that we compared (21 in total), we sorted the results of all statistical tests in ascending order of $p$-value, and adjusted the significance level, $\alpha_{\textrm{sig}}$, for these tests according to the Holm-Bonferroni method (\citealt{Holm1979}).      

\subsection*{Characteristics of individuals' interactions}

After application of a Holm-Bonferroni correction the only parameter that differed between fish that spent the greatest proportion of time at the front of their pair and their partner was mean change in speed over time -- a parameter intrinsically linked with how an animal interacts with its neighbours (\citealt{HerbertReadetal2011,Katzetal2011}). We therefore analysed in more detail how individuals that were more often observed at the front of a pair responded differently to their partner's position and movements, compared to fish that were more often observed at the back of the pair.

%%%%%%%%%%%%%%%%%%%% Figure %%%%%%%%%%%%%%%%%%%%%%
\begin{figure*}
\centering\includegraphics[width=\textwidth]{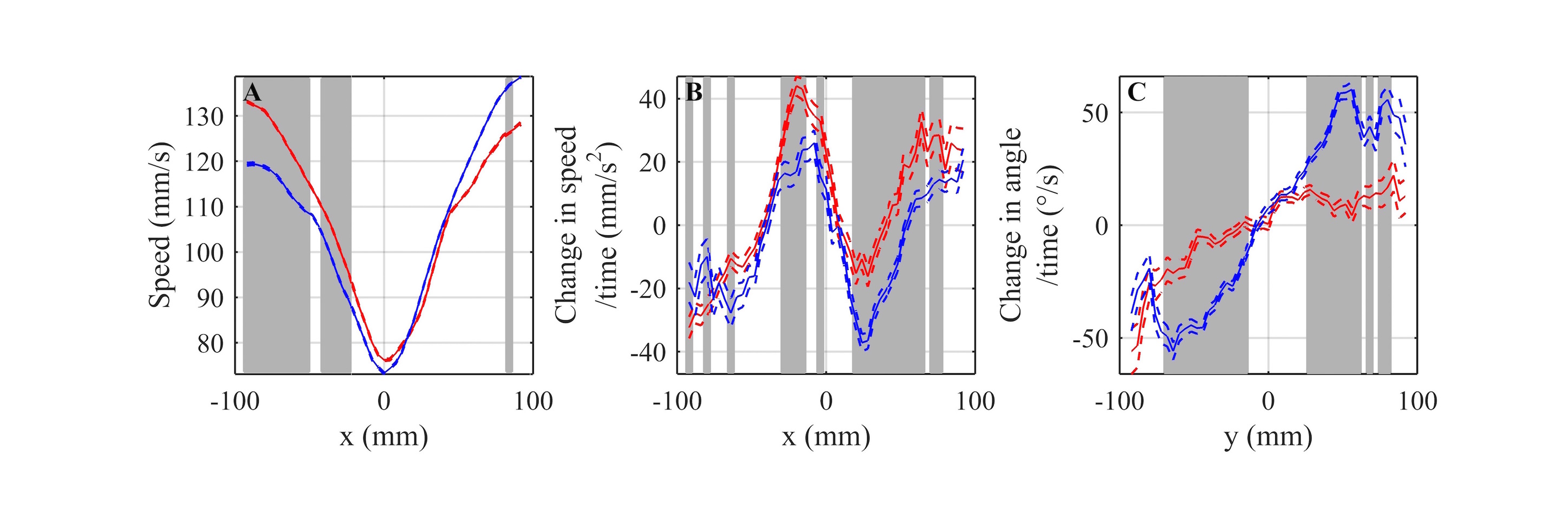}
\caption{A The mean speed, B mean change in speed over time, C mean change in angle of motion over time of front fish (solid red curve) and back fish (solid blue curve) as a function of the relative $x$-coordinate of their partner (A \& B) or right-left position (C). Dashed curves are plotted one standard error above and below all means in each panel. Combined, the mean change in speed over time (B) and the mean change in angle of motion over time (C) describe how an individual adjusts its velocity as a function of relative partner location. Grey regions in all plots highlight where our randomisation procedure indicated that the sign and magnitude of the difference between the front fish curves and back fish curves was unlikely to occur if fish had randomly been allocated to the sets of front or back fish.}
\label{fig:Fig2}
\end{figure*}
%%%%%%%%%%%%%%%%%%%% Figure %%%%%%%%%%%%%%%%%%%%%%

We first produced heat maps to illustrate where individuals were most commonly observed relative to a focal individual's position and direction of motion. We produced plots showing how an individual's speed, change in speed over time and change in heading over time changed as a function of the relative $(x,y)$ position of their partner, as well as a function of $x$ (front:back) or $y$ (left:right) only (see; \citealt{Katzetal2011}). Again, we compared these separately for front fish reacting to the position of back fish, and back fish reacting to the position of front fish. We performed a series of calculations to estimate the probability that observed differences in data projected onto the $x$- or $y$-axes could result if fish were randomly allocated to the two categories (front fish or back fish). Details of the method we used to construct our heat maps, the projection of these maps onto $x$- and $y$-axes and our subsequent randomisation analysis are provided in online appendix A.

To determine if there were any differences in the pair's behaviour when individuals occupied different positions in the pair, we compared the median speed of the pair's centroid when the front fish was ahead of its partner, versus times when the front fish was behind its partner. This measure gave an estimate of the exploration rate of the pair when in different spatial configurations (see online appendix A).

\section*{Results}

Fish tended to maintain positions in front or behind their partner (fig.~\ref{fig:Fig1}). The majority of encounters between the pair occurred when fish were separated by approximately 25 mm to 90 mm (in the front:back $x$-direction). When a fish was at the front of the pair, it directed group movement in the majority of cases: the time-lag associated with maximum directional correlation was positive for 73 out of 80 fish when in the front-most position ($p = 5.7841 \times 10^{-15}$, two-tailed binomial test). Individuals led their partner, therefore, when at the front of the group. The proportion of frames that the front fish was found at the front of the group ranged widely from 0.5004 to 0.9643 (see Tables A1 and A2 and fig. A1 in online appendix A).

Fish that had an additional 5 minutes to explore the arena were not significantly associated with being the front fish ($p=0.4119$, N = 20, n = 9, two-tailed binomial test). After application of the Holm-Bonferroni correction, the only difference in movement parameters between front fish and back fish was the mean change in speed over time; front fish had greater mean changes speed over time than back fish (see Table A5 in online appendix A for details of statistical tests applied to movement parameters and body length). 

We then tested whether the movements of front fish or back fish differed in response to their partner's position. Speeds of both the front fish and the back fish were lowest when their partner was close to them ($x=0$). The average speeds adopted by front fish were approximately symmetric about $x=0$ (fig.~\ref{fig:Fig2} A); front fish tended to travel at a similar speed regardless of whether their partner was at the same distance in front or behind them (for example, compare the red curve speeds when $x=50$ mm with $x=-50$ mm in fig.~\ref{fig:Fig2} A). In contrast, there was more pronounced asymmetry in the speeds of back fish as a function of their neighbour's position. Back fish tended to adopt lower speeds when their partner was located behind them, and higher speeds when their partner was located in front of them (fig.~\ref{fig:Fig2} A). The differences in the speeds adopted by front or back fish when their partner was behind them ($x<-25$ mm) were unlikely to occur as a result of randomly allocating fish to categories of front fish or back fish (see fig. A16 E). 

%%%%%%%%%%%%%%%%%%%% Figure %%%%%%%%%%%%%%%%%%%%%%
\begin{figure*}
\includegraphics[width=\textwidth]{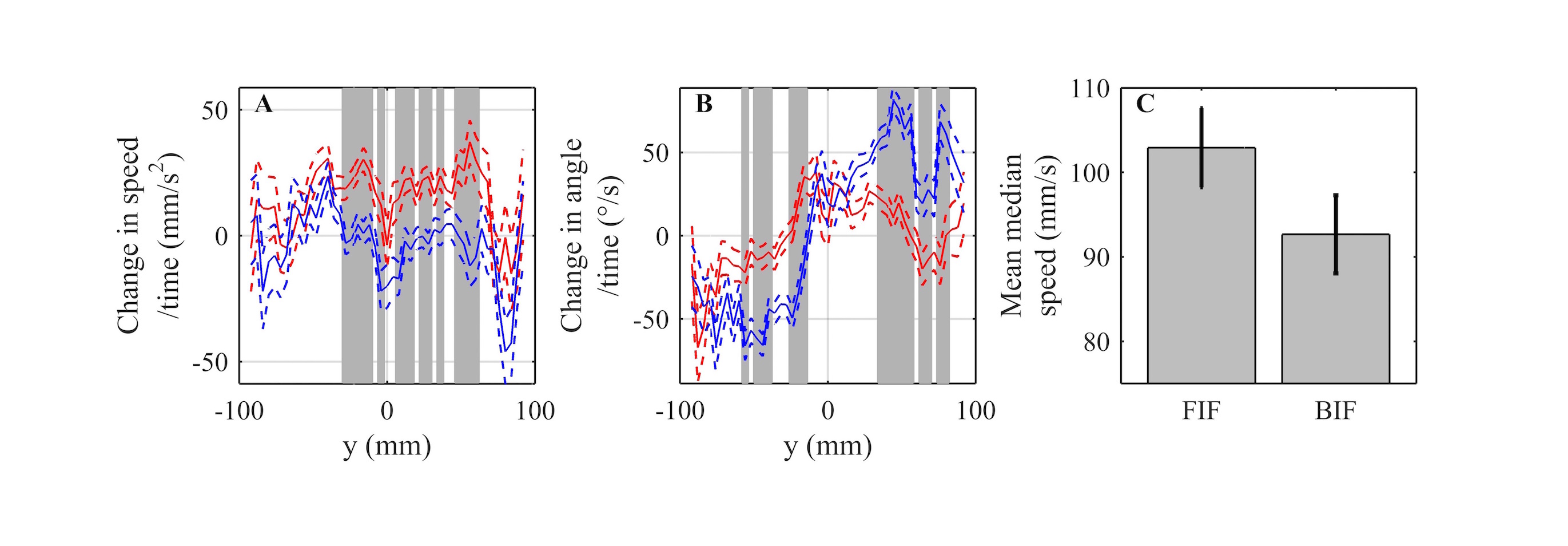}
 \caption{The mean change in speed over time (A) and the mean change in angle of motion over time (B) of front fish (solid red curves) and back fish (solid blue curves) as a function of the relative $y$-coordinate of their partner when both fish were approximately side by side (such that the relative $x$-coordinate of a partner fish satisfied $-32 < x \leq 32$ mm). Dashed curves are plotted one standard error above and below the means. Panel C illustrates the mean (across all pairs of fish; $\pm 1$ SE) median speed of the centroid of each pair of fish across frames where front fish (left bar) or back fish (right bar) were located at the front of each pair (FIF = Front fish In Front, BIF = Back fish In Front). Grey regions in A and B highlight where our randomisation procedure indicated that the sign and magnitude of the difference between the front fish curves and back fish curves was unlikely to occur through random categorisation of fish.}
\label{fig:Fig3}
\end{figure*}
%%%%%%%%%%%%%%%%%%%% Figure %%%%%%%%%%%%%%%%%%%%%%

Fish also adjusted their speed depending on their partner's position. The instantaneous changes in speed over time as a function of their partner's position were qualitatively similar for both front fish and back fish. If a neighbour was located a short distance behind the focal fish (in the domain $-50 \leq x \leq 0, \, -50 \leq y \leq 50$ (mm)), then the focal fish would increase its speed, acting to move away from its partner (fig.~\ref{fig:Fig2} B and fig. A18, A \& B).  However in this region, this change in speed was higher for front fish reacting to the position of the back fish (fig. \ref{fig:Fig2}B, red curve), than for back fish reacting to position of the front fish (fig.~\ref{fig:Fig2} B, blue curve). If a neighbour was located a short distance in front of the focal fish (within the domain $0 \leq x \leq 50, \, -50 \leq y \leq 50$ (mm)), then the focal fish would decrease its speed (again see fig.~\ref{fig:Fig2} B and fig. A18, A \& B). Once again, the mean change in speed in this region was greater for front fish reacting to back fish than for back fish reacting to front fish. In fact, the front fish's mean change in speed projected onto the $x$-axis over the range $-50 \leq x \leq 50$ (mm) was always greater than the back fish's mean change in speed. The magnitude in difference between the front fish's and back fish's mean change in speed ranged from close to 0 mm/s$^2$ to over 20 mm/s$^2$. Our randomisation analysis suggested that these observed differences in the intervals $-25 \leq x \leq -10$ (mm), $10 \leq x \leq 60$ (mm) and $-10 \leq y \leq 35$ (mm) would be statistically unlikely if fish were randomly identified as front fish or back fish (figs. A19, E \& F). Hence when their partner was close to them, front fish or back fish adjusted their speed differently as a function of relative partner location.

A fish may come to occupy the front position in a pair in multiple ways that include: swimming faster than its neighbour when they are side by side (active overtaking), maintaining its speed whilst its neighbour slows down (passive overtaking), or through these mechanisms acting at the same time.  To investigate the mechanism behind individuals adopting front positions, we compared each fish's change in speed over time (projected onto the $y$-axis) when both fish were approximately side by side (where $-32 < x \leq 32$ (mm), that is, where the difference in the $x$-coordinates of the fish differed by up to a little over one body length). Front fish generally had a greater mean change in speed over time than their partner in this region (fig.~\ref{fig:Fig3} A). Further, front fish tended to exhibit positive changes in speed, whereas their partners exhibited changes in speed that were closer to zero, and in some instances negative (fig.~\ref{fig:Fig3} A), especially over the approximate range $-25 \leq y \leq 60$ mm. These results suggest that front fish tended to accelerate to either take or maintain the frontmost position (active overtaking), whereas back fish tended to continue at their current speed or even reduce their speed when the front position was in contest, in effect, giving way to their partner. Our randomisation analysis suggested that the sign and magnitude of the difference in leader and follower changes in speed over time in fig.~\ref{fig:Fig3} A were unlikely to occur if fish were randomly allocated to front fish or back fish categories for data in smaller regions contained within the range $-30 \leq y \leq 60$ mm (see fig. A22, E).

In general, both front fish and back fish exhibited a tendency to turn towards their partners, with fish turning anti-clockwise (characterised by positive changes in angle of motion) when their partner was to their left (positive $y$), and clockwise when their partner was to their right (characterised by negative changes in angle of motion for negative $y$) (fig.~\ref{fig:Fig2} C and fig. A20, A \& B). The magnitude of mean changes in angle of motion  as a function of $y$ tended to be larger for back fish turning towards front fish (blue curve) than for front fish turning towards back fish (red curve) (fig.~\ref{fig:Fig2}C). Such a difference was unlikely to occur through random allocation of fish to the categories of front fish or back fish over the approximate ranges $-60 \leq y \leq -10$ (mm) and $20 \leq y \leq 60$ (mm) (fig. A21, F). When fish were approximately side by side (where $-32 < x \leq 32$ (mm), these patterns remained, with back fish tending to turn towards their partner with greater turning speed than front fish (fig.~\ref{fig:Fig3} B). 

The median speed of the centroid of the pair differed when either front fish or back fish were in front ($p = 0.0027$, $W = 633$, $z=2.9974$, two-sided Wilcoxon paired-sample test, median difference in median speeds = 7.3128 mm/s); in general pairs moved at greater median speeds when front fish were in front versus when back fish were in front (fig.~ \ref{fig:Fig3} C).

\section*{Discussion}

Whilst informational state could not predict whether an individual would occupy the front of the group for the greatest proportion of time, we found that fish that dominated the front position exhibited the greatest mean change in speed over time. Fish who occupied the front or back position in the pair for the majority of time differed in how they adjusted their velocity based on their partner's relative position. At a group level, pairs tended to travel at greater speeds when the fish that most often occupied the front position was at the front versus when the fish that tended to occupy the back position was at the front.    

Our results suggest two mechanisms that explain why individuals responded to their partner's position differentially. First, the rate of turning towards their partners location suggests individuals may differ in their likelihood to copy or follow the movements of others. Individuals more often observed at the back of groups had higher turning rates to orientate towards their partner's position than individuals more often observed at the front of groups.  Back fish were also slower when they found themselves ahead of their partner than when they were behind their partner. These patterns suggest higher degrees of social responsiveness of the back fish than the front fish. The reduced tendencies of the the front fish to turn towards its partner's position and its symmetrical speed distribution as a function of its partner's anterior-posterior position suggest these individuals were less responsive to the their partner's position. Indeed, leadership through social indifference has been explored theoretically (\citealt{conradt2009leading}). More recently, it has been demonstrated by training individual fish that in order to be effective leaders, individuals should balance their own goal oriented behaviour with how responsive they are to their neighbours (\citealt{ioannou2015potential}). Doing so acts to maintain group cohesion, whilst also allowing information to propagate through the group allowing individuals to lead others (\citealt{ioannou2015potential}). Differences in social responsiveness, therefore, can have large-scale implications for group dynamics.  

Second, the rules of interaction we have identified here and previously (\citealt{HerbertReadetal2011}) reveal equivalents to zones of repulsion, or the effective range of similar avoidance terms seen in many models of collective motion (\citealt{Couzinetal2002,Couzin2005,Janson2006,Diwold2011,FetecauandGuo2012}). However, they may also be interpreted as responses that enable individuals to occupy particular spatial positions in the group (\citealt{perna2014note}). We find partial evidence for this second interpretation in our data.  When fish were approximately side by side, the front fish had, on average, positive changes in speed over time. Back fish, on the other hand, had lower, and close to zero or negative changes in speed when their partner was beside them. Such an effect could be interpreted as front fish acting to move to positions in front of their partners, whilst back fish being less driven to occupy those positions. A combination of social responsiveness and positional preferences, therefore, are likely to determine the positioning behaviour of individuals in pairs of shoaling mosquitofish.  

Why might fish differ in their responsiveness to neighbours or their willingness to occupy particular spatial positions? Slight differences in internal nutritional state, or differences in aerobic scope between individuals, could drive these differences (\citealt{killen2011aerobic}). Whilst we fed fish at the same time, there is the possibility that individuals differed in their metabolic rates leading to different energetic requirements. If the less satiated individual initiated more attempts to explore their environment to find food, then such differences could manifest in reduced social responsiveness to their partner. Hungry individuals or individuals with poorer body condition sometimes lead partners more often than satiated individuals or individuals with a higher quality body condition (\citealt{nakayama2012initiative,ost2013relative}). Generally, hungry fish tend to move closer to the front of larger groups and their satiated counterparts tend to occupy positions closer to the rear (\citealt{Krauseetal1992,Krause1993,Hansenetal2015b}). Such a mechanism is consistent with theoretical studies where leadership can spontaneously emerge as a result of intrinsic state differences between individuals (\citealt{rands2003spontaneous}). Moreover, internal nutritional state (hunger or satiation) has been shown to have an effect on the basic locomotion of mosquitofish (\citealt{Hansenetal2015a}). Mosquitofish left unfed for a period of 24 hours moved with greater mean speed than those that had been fed to satiation, but unlike fish that occupied the front of pairs in the experiments described here, hungry mosquitofish also tended to exhibit greater median turning speeds (\citealt{Hansenetal2015a}). 

Whilst state and body size may be important determinants of leadership, other differences, such as an individuals' personality may also result in increased or decreased social responsiveness (\citealt{harcourt2009social}). Theoretical predictions suggest that intrinsic differences between individuals in their social responsiveness can be maintained through frequency dependent selection, acting to stabilise the roles of `leaders' and `followers' in populations (\citealt{johnstone2011evolution}). However, whether such roles exist in large fission-fusion systems remains to be empirically tested.  We found a wide degree of variation in how often individuals occupied positions at the front of the group (50.04\% to 96.43\%) and this may be explained by the difference in social responsiveness between the two individuals in the pair. If individuals occupy similar social responsiveness levels, this may lead to sharing of leadership roles through `turn-taking' strategies (\citealt{harcourt2010pairs}). On the other hand, individuals that have disparate levels of social responsiveness may simply adopt the role in the pair that matches their level of responsiveness. Whether the adoption of behavioural roles in groups has some functional benefit for individuals in groups remains to be investigated further. Here, when individuals occupied positions that they were most commonly observed in (i.e. front fish in front and back fish at the back of the pair) the pair explored their environment more quickly. This suggests that if individuals can gauge their relative roles in group, either through passive self-assortment or actively adjusting their behaviour to suit their partner's social phenotype, the pair may collectively realise the benefits of group living by remaining cohesive, whilst also exploring their environment more quickly (\citealt{krause2002liv}). Whether social responsiveness is a consistent and heritable trait in an individual's behaviour should be investigated with future selection experiments and repeated tests on individuals.

Our results highlight that individuals differ in their responses and movements to their neighbours and these differences can give rise to differential spatial positioning in free moving groups. A further exploration of the consistency and variability of these responses should now be made in detail. The ability to identify differences in the movement responses between individuals using their movement trajectories now allows us the opportunity to investigate the evolution and maintenance of responsive types in natural populations.  

%%%%%%%%%%%%%%%%%%%%%
% Acknowledgments
%%%%%%%%%%%%%%%%%%%%%
% You may wish to remove the Acknowledgments section while your paper 
% is under review (unless you wish to waive your anonymity under
% double-blind review) if the Acknowledgments reveal your identity.
% If you remove this section, you will need to add it back in to your
% final files after acceptance.

\section*{Acknowledgments}

We thank Andrea Perna for useful suggestions and comments on the manuscript. The authors acknowledge that this work was funded by the Australian Research Council via DP130101670 and by the Knut and Alice Wallenberg foundation grant: 102 2013.0072.

\bibliographystyle{amnat}
\bibliography{ARCreferences_Edit}

\end{multicols}

\clearpage
\newpage{}

\renewcommand{\thesection}{\Alph{section}}
\renewcommand\thefigure{\thesection A\arabic{figure}}    
\setcounter{figure}{0}  
\setcounter{table}{0}
\renewcommand{\thetable}{A\arabic{table}}

\section*{Online Appendix A: Supplementary Methods}

\subsection*{Basic individual characteristics}\label{supp:BasicCharacteristcs}
We determined each fish's velocity, speed, change in speed over time, acceleration, magnitude of acceleration, turning speed and body length directly from tracking data using the following series of calculations. 

Writing $(x_{i}(t),y_{i}(t))$ as the coordinates of fish $i$ at time $t$ we determined the $x$ and $y$ components of a fish's velocity using the standard forward-difference approximations:
\begin{equation}
u_{i}(t) = \frac{x_{i}(t + \Delta t) - x_{i}(t)}{\Delta t} \qquad \textrm{and} \qquad v_{i}(t) = \frac{y_{i}(t + \Delta t)-y_{i}(t)}{\Delta t},\label{eq:individualvelocity}
\end{equation}
where $\Delta t = 1/40$ s was the constant duration between consecutive video frames. A fish's speed at time $t$ was then approximated as:
\begin{equation}
s_{i}(t) = \sqrt{(u_{i}(t))^2 + (v_{i}(t))^2}.\label{eq:individualspeed}
\end{equation}
Following immediately from this calculation we determined the change in a fish's speed over time via:
\begin{equation}
\frac{\Delta s_{i}}{\Delta t}(t) = \frac{s_{i}(t + \Delta t)-s_{i}(t)}{\Delta t}.\label{eq:tangentialacceleration}
\end{equation}
(The above measure is referred to as tangential acceleration in \citealt{HerbertReadetal2011}.) The measure in equation (\ref{eq:tangentialacceleration}) differs from both the acceleration of a fish (a vector), and the magnitude of acceleration. $\frac{\Delta s}{\Delta t}$ can take negative values (representing deceleration), so it is more illuminating to examine than magnitude of acceleration (which is non-negative by definition) when it is of interest to determine when fish are speeding up or slowing down. 

We determined the $x$ and $y$ components of a fish's acceleration respectively using the centred difference approximations:
\begin{equation}
b_{i}(t)=\frac{x_{i}(t + \Delta t) - 2x_{i}(t)+x_{i}(t - \Delta t)}{(\Delta t)^2} \qquad \textrm{and} \qquad c_{i}(t)=\frac{y_{i}(t + \Delta t) - 2y_{i}(t)+y_{i}(t - \Delta t)}{(\Delta t)^2},\label{eq:individualacceleration}
\end{equation}
and thus the magnitude of a fish's acceleration was determined by:
\begin{equation}
a_{i}(t)=\sqrt{(b_{i}(t))^2+(c_{i}(t))^2}.\label{eq:magofacceleration}
\end{equation} 
 
We estimated a fish's turning speed at time $t$ based on the direction of its velocity vector at times $t$ and $t + \Delta t$. To do this we constructed unit vectors in the direction of each fish's velocity vector, with components:
\begin{equation}
\hat{u}_{i}(t)=\frac{u_{i}(t)}{s_{i}(t)} \qquad \textrm{and} \qquad \hat{v}_{i}(t)=\frac{v_{i}(t)}{s_{i}(t)}.\label{eq:unitvelocity}
\end{equation}
The internal angle between the unit vectors for a given fish's direction of motion at at times $t$ and $t + \Delta t$ was then determined using the dot product; we then divided this angle by the duration between consecutive frames to estimate turning speed. Compactly, the formula for calculating a fish's turning speed (in degrees/s) can be written as:
\begin{equation}
\alpha_{i}(t)=\frac{180}{\pi}\frac{\cos^{-1}(\hat{u}_{i}(t)\hat{u}_{i}(t + \Delta t)+\hat{v}_{i}(t)\hat{v}_{i}(t + \Delta t))}{\Delta t}. \label{eq:turningspeed}
\end{equation}  

Included in Ctrax output are measurements of the major and minor axes of an ellipse that is fitted to the image of each individual for each video frame (in practice the output measurements are one quarter of the length of the major and minor axes). We used the median size of the major axis as an estimate of each fish's body length $(L)$.

\subsection*{Within group position}\label{sssec:Leadership_group_position}
We determined the ordering of the pair of fish relative to the direction of motion of the group centre using the following calculations and linear transformations. For each video frame we identified the mean coordinates of the pair of fish $(\bar{x}(t),\bar{y}(t))$ (that is, the group centre). We then estimated the velocity of the group centre at time $t$ using:
\begin{equation}
u_{c}(t) = \frac{\bar{x}(t + \Delta t)-\bar{x}(t)}{\Delta t} \quad \textrm{and} \quad v_{c}(t) = \frac{\bar{y}(t + \Delta t)-\bar{y}(t)}{\Delta t}.\label{eq:groupcentrevelocity}
\end{equation}
Next, for each time step, we shifted the coordinates of each fish so that the origin of the coordinate system lay at the group centroid, and then rotated the coordinates of the fish so that the direction of motion of the group centre (derived from equation (\ref{eq:groupcentrevelocity})) was parallel to and pointed in the same direction as the positive $x$-axis. The transformed coordinates of the fish meant that the fish with the greatest $x$-coordinate was at the front of the pair for a given frame. We counted the number of frames that each fish was located in the forward-most position of the pair; we then identified the individual that spent the greatest proportion of frames at the front of the pair (which we term the `front fish' for brevity) and hence the individual that spent the greatest proportion of frames at the back of the pair (termed the `back fish'). In general, individuals swapped positions throughout most experiments even though one individual was more often found at the front of the group (fig.~\ref{fig:totalprop_hist}).

To examine whether the group properties changed depending on whether different individuals occupied different positions in the shoal, we calculated the speed of the group centre:
\begin{equation}
s_{c}(t) = \sqrt{\left(u_{c}(t)\right)^2+ \left(v_{c}(t)\right)^2}.\label{eq:centroidspeed}
\end{equation}
We then identified all frames where the front fish was at the front of the pair, and determined the median speed of the group centre across these frames (for each pair). Similarly, we identified all frames where the back fish was in front, and determined the median speed of the group centre across these frames. We treated the median speeds when the front fish or back fish was in front as paired samples, and performed a two-sided Wilcoxon paired-sample test (see for example, \citealt{Zar1996}) to determine if the group's median speed differed when front fish or back fish were in front.     

\subsection*{Directional correlation and delay associated with maximum directional correlation}\label{sssec:Leadership_timelag}
It is not always the case that an animal located at the front of a group is responsible for directing the motion of the group. For example, streaker bees guide honey bee swarms by flying rapidly through the upper portions of a swarm from the rear to the front of the group (\citealt{Beekman2006,Janson2006,Schultz2008,Latty2009,Diwold2011}). To examine whether individuals guided the motion of their partner when they were at the front of the group, we examined the directional correlation of the fish (see for example \citealt{Nagy2010,Katzetal2011}). Using equation (\ref{eq:unitvelocity}) we obtained each fish's direction of motion in component form for all time-steps, $\left(\hat{u}_{i}(t),\hat{v}_{i}(t)\right)$. We identified the fish in each trial as fish 1 or 2 based on the order in which their trajectories were recorded, and then produced two sets of time series of each fish's direction of motion. The first of these time series left all entries where fish 2 was in the frontmost position blank (that is, it only contained information for the time steps when fish 1 was in front), and the second time series left all entries where fish 1 was in front blank. For each set of time series (corresponding to fish 1 in front or fish 2 in front), we determined the directional correlation:
\begin{equation}
C_{ij}(\tau)=\left<\hat{u}_{i}(t)\hat{u}_{j}(t + \tau) +\hat{v}_{i}(t)\hat{v}_{j}(t + \tau)\right>, \label{eq:directionalcorrelationdelay}
\end{equation}    
where $\tau = \tau_{n} \Delta t$ is time-lag in seconds, $\tau_{n} \in \left\{-120,-119,\ldots,120\right\}$ is the number of frames corresponding to a given time-lag and $\left< \cdot \right>$ represents the mean taken over all $t$. (The term inside the angle brackets is the dot/inner product of the direction of motion of fish $i$ at time $t$ and the direction of motion of fish $j$ at time $t + \tau$.) We then identified the maximum value of $C_{ij}(\tau)$ and the value of $\tau$ that corresponds to this maximum, denoted $\tau^{*}_{ij}$. Provided that $C_{ij}(\tau^{*}_{ij})$ was large enough to suggest that there was reasonable correlation in the directions of motion of the two fish, a positive value of $\tau^{*}_{ij}$ suggested that fish $j$ adjusted its direction of motion to match that adopted by fish $i$ at an earlier time (that is, fish $j$ was following the direction of fish $i$), whereas a negative value of $\tau^{*}_{ij}$ suggested that fish $i$ was following fish $j$.

\subsection*{Characteristics of interaction}\label{supp:RulesofInteraction}
The first step in making each heat map was to determine the distance between the pair of fish for all times $t$:
\begin{equation}
d(t) = \sqrt{(x_{2}(t)-x_{1}(t))^{2}+(y_{2}(t)-y_{1}(t))^{2}}.\label{eq:iid}
\end{equation}
Next we calculated the angle between the direction of motion of each fish, $i$, (given in component form by equation (\ref{eq:unitvelocity})) and the directed straight line segment from the location of fish $i$ to the location of its partner, fish $j$, for all $t$. To aid in this calculation, we constructed a unit vector in the direction of the straight line segment from fish $i$ to fish $j$, with components:
\begin{equation}
\hat{x}_{ij}(t) = \frac{x_{j}(t)-x_{i}(t)}{d(t)} \qquad \textrm{and} \qquad \hat{y}_{ij}(t) = \frac{y_{j}(t)-y_{i}(t)}{d(t)}. \label{eq:unit_i_to_j}
\end{equation}
The internal angle between the unit vectors representing the direction of motion of fish $i$ (equation (\ref{eq:unitvelocity})) and the direction from fish $i$ to fish $j$ (equation (\ref{eq:unit_i_to_j})) can be determined using a dot product (similar to equation (\ref{eq:turningspeed})):
\begin{equation}
\phi_{ij}(t) = =\frac{180}{\pi}\cos^{-1}(\hat{u}_{i}(t)\hat{x}_{ij}(t)+\hat{v}_{i}(t)\hat{y}_{ij}(t)). \label{eq:phi_ij}
\end{equation}
Using equation (\ref{eq:phi_ij}) will determine an angle constrained so that $0 \leq \phi_{ij} \leq 180^{\circ}$. An additional calculation is required to determine if fish $j$ is either to the left or the right of fish $i$. Relative to the direction of motion of fish $i$, fish $j$ lies to the left (right) of fish $i$ if the sign of the following equation is positive (negative):
\begin{equation}
\lambda_{ij}(t)=\textrm{sgn}\left(\hat{u}_{i}(t)\hat{y}_{ij}(t) - \hat{v}_{i}(t)\hat{x}_{ij}(t)\right).\label{eq:right_left_sign}
\end{equation} 
The term in the parentheses of equation (\ref{eq:right_left_sign}) is the vertical component of the cross-product of the unit vector pointing in the direction of motion of fish $i$ with the unit vector pointing from fish $i$ to fish $j$. We defined the signed angle between the direction of motion of fish $i$ and the relative location of fish $j$ as:
\begin{displaymath}
\varphi_{ij}(t)=\left\{ \begin{array}{ll}
		    \lambda_{ij}(t)\phi_{ij}(t) & \textrm{if $\lambda_{ij}(t) \neq 0$,}\\
	               \phi_{ij}(t) & \textrm{if $\lambda_{ij}(t) = 0$.}
	               \end{array} \right.
\end{displaymath}

Each heat map was constructed in Cartesian coordinates $(x,y)$, where $-100 \leq x \leq 100$ (mm) and $-100 < y \leq 100$. Focal fish were treated as being located at the origin, moving to the right (parallel to the $x$-axis). A separate map was produced for the sets of fish that spent the greatest proportion of time at the front of their pair and fish that spent the greatest proportion of time at the back of their pair for each quantity of interest. (Here we discuss calculations relating to speed by means of example, but the method is identical for other quantities.)  

We converted the relative locations of partner fish from the polar form described by $(d(t), \varphi_{ij}(t))$ to Cartesian coordinates via:
\begin{eqnarray}
x_{ij,\textrm{relative}}(t) & = & d(t) \cos \left( \varphi_{ij}(t)\right),\\
y_{ij,\textrm{relative}}(t) & = & d(t) \sin \left( \varphi_{ij}(t)\right).
\end{eqnarray}

We divided the domain centred on each focal fish into a set of overlapping bins such that the left edges of the bins were located at $x_{l,\textrm{left}}=-100, -96, -92, -88, \ldots, 84$ (mm), the right edges of the bins were located at $x_{l,\textrm{right}}=-84, -80, -76, -72, \ldots, 100$ (mm), the bottom edges of the bins were located at $y_{k,\textrm{bottom}}=-100, -96, -92, -88, \ldots, 84$ (mm) and the top edges of the bins were located at $y_{k,\textrm{top}}=-84, -80, -76, -72, \ldots, 100$ (mm). That is, bins extend 16 mm in both the $x$ and $y$ directions (approximately half a body length), and were separated by 4 mm in both $x$ and $y$ directions. The biological reason behind using such smoothing is that it is reasonable to assume that small changes in the relative position of partner fish should not result in dramatically different behaviour of focal fish (on average). 

For each fish $i$ in a given set, and each time-step, fish $i$'s speed at time $t$ was included in bin $(l,k)$ if  $x_{l,\textrm{left}} <x_{ij,\textrm{relative}}(t) \leq x_{l,\textrm{right}}$ and $y_{k,\textrm{bottom}} <y_{ij,\textrm{relative}}(t) \leq y_{k,\textrm{top}}$. Once data corresponding to all fish and time steps were allocated to bins, we calculated the mean of the finite entries in each bin, and rendered the results with the help of MATLAB's intrinsic \emph{surf} function. In the case where alignment was the quantity of interest, we determined the mean angle between the facing direction of the focal fish and their partners using standard methods of circular statistics \citealt{Zar1996} (plotted as arrows in the relevant plots), along with $R$, which is a measure of the scatter of all the angles in a set. For reference, the mean, $\bar{\vartheta}$, of a set of angles, $\vartheta_{i}$, is given by:
\begin{equation}
\bar{\vartheta}=\tan^{-1}\left(\frac{Y}{X}\right),\label{eq:meanangle}
\end{equation}
where $X=\sum_{i=1}^{n}\cos \vartheta_{i}$, $Y=\sum_{i=1}^{n}\sin \vartheta_{i}$, and      
\begin{equation}
R=\frac{\sqrt{X^2+Y^2}}{n}.\label{eqn:rsq}
\end{equation}
In surface plots of alignment, colours corresponded to the $R$ value in each bin, rather than a mean.

In addition to the magnitude of turning speed given by equation (\ref{eq:turningspeed}), we required information about the sense of rotation of fish (clockwise or anti-clockwise) to construct appropriate plots of turning behaviour. This sense of rotation was determined by examining the vertical component of the cross product of unit velocity vectors for a fish at times $t$ and $t + \Delta t$, similar to the calculations for $\lambda_{ij}(t)$ in equation (\ref{eq:right_left_sign}). We refer to the quantity that combines sense of rotation and magnitude of turning speed as change in angle of motion over time, change in angle over time or change in heading, denoted $\frac{\Delta \theta}{\Delta t}$.  

In addition to surface plots, we produced line-graphs of the proportion of encounters with neighbour fish, mean speed of focal fish, mean change in speed over time of focal fish and mean change in angle of motion over time of focal fish by projecting data contained in the square bins (described above) onto both the $x$ and $y$-axes. Data was projected onto the $x$-axis by combining all data that satisfied $x_{l,\textrm{left}} <x_{ij,\textrm{relative}}(t) \leq x_{l,\textrm{right}}$ into bin $l$, irrespective of the $y$-coordinate associated with each data point. Similarly, data was projected onto the $y$-axis by combining all elements that satisfied $y_{k,\textrm{bottom}} <y_{ij,\textrm{relative}}(t) \leq y_{k,\textrm{top}}$ into bin $k$. As well as calculating means for the line-graphs, we determined the standard deviation of values contained in each bin, and hence standard errors (based on a sample size equal to the number of elements contained in a given bin). Denoting curves associated with fish that spent the greatest proportion of time at the front of their pair as $A(x)$ (or $A(y)$) and curves associated with fish that spent the greatest proportion of time at the back of their pair as $B(x)$ (or $B(y)$), we determined the difference in the proportions or means associated with each line graph ($A(x) - B(x)$) for subsequent analysis.

We were interested in how fish that occupied the front or back position most frequently adjusted their velocity on average when their partners were approximately beside them. To examine this behaviour we projected our data for change in speed over time and change in angle of motion/heading over time onto the $y$-axis using the method outlined in the previous paragraph, but only using data that satisfied the condition $-32 < x \leq 32$ (mm) -- a range that corresponds to the fish being approximately side by side (up to a difference in centres of a little over one body length, see fig. \ref{fig:bodylengths_boxplot}), and potentially contesting the frontmost position of the pair. 

We performed two sets of additional calculations where one fish in a pair was randomly allocated to the set of front fish with probability 0.5 and their partner was allocated to the set of back fish, in an attempt to examine the likelihood that any of the trends that appeared in our line-graphs could arise from random categorisation of each fish rather than a tendency to occupy a given position. One set was comprised of 100 random allocation processes where all forty pairs of fish had one member randomly allocated to the pool of front fish and the other member allocated to the back fish pool; the other set was comprised of 1000 random allocation processes performed in the same manner. For each randomisation, we first produced line-graphs of each quantity (as described above) for the set of randomly selected `front' and `back' dominant fish. From these graphs we determined the differences obtained from curves for the randomly allocated sets of front fish minus the randomly allocated sets of back fish (denoted $A_{r,n}(x) - B_{r,n}(x)$) for the $n$th random allocation). Once all curves from all randomisations were determined and stored, we estimated the probability that a difference in curves would have the same sign as that observed, and that the magnitude of that difference was at least as big as that observed via a count. To do this, for each bin associated with each curve we counted all the instances where the differences in randomised front fish and back fish curves ($A_{r,n}(x) - B_{r,n}(x)$) were greater than or equal to that observed from position based identification of front fish and back fish ($A(x) - B(x)$) when the position based differences were positive, and all the instances where $A_{r,n}(x) - B_{r,n}(x) \leq A(x) - B(x)$ when $A(x) - B(x) < 0$. Finally, when we examined projections of change in speed over time and change in angle of motion over time onto the $y$-axis for the smaller range over $x$ of $-32 < x \leq 32$ (mm), our associated probabilities were derived from a set of 100 random allocation processes only (examination of the probabilities derived from 100 and 1000 random allocation process for other projections indicated that the tenfold increase in random allocations did not have a major effect on the estimated probability). 

\section*{Online Appendix A: Supplementary Results}
\subsection*{Proportion of time at group front for each fish}\label{supp:proportionoftimetables}
\begin{figure}[!h]
	\centering
	\includegraphics[width=\textwidth]{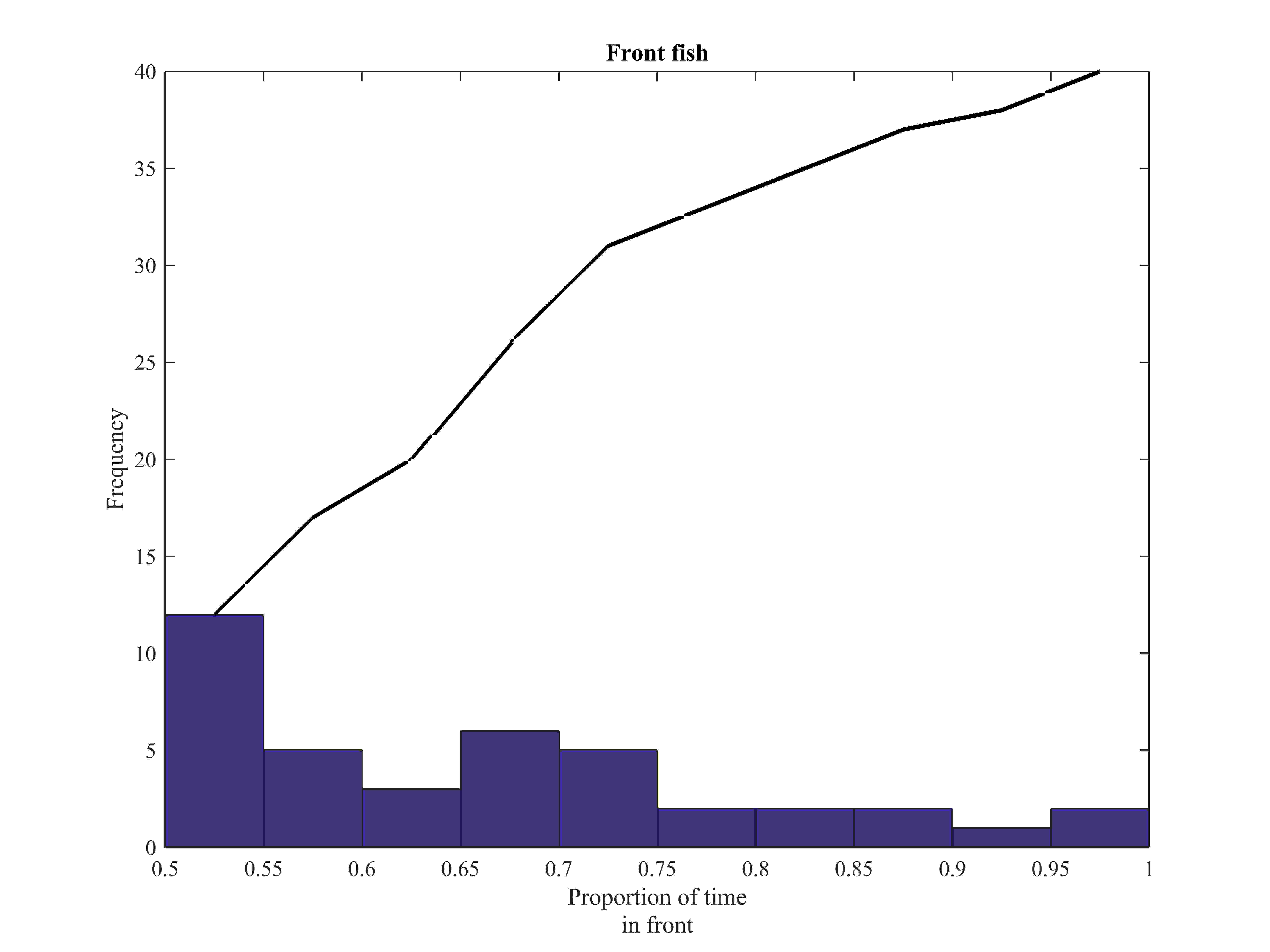}
	%\end{center}
	\caption{Total proportion of time that the front fish spent in front of their partner relative to the group centroid (when the fish were separated by 100 mm or less). The black line shows the cumulative sum of this distribution (number of fish).}\label{fig:totalprop_hist}
\end{figure}

Tables \ref{tab:proportions1} and \ref{tab:proportions2} list the proportion of time spent at the front of the group by each fish in each trial. Figure \ref{fig:totalprop_hist} contains a frequency histogram of the total proportion of time spent at the front of each pair by `front' fish.

\begin{figure}[!h]
	\begin{center}
	\includegraphics[width=\textwidth]{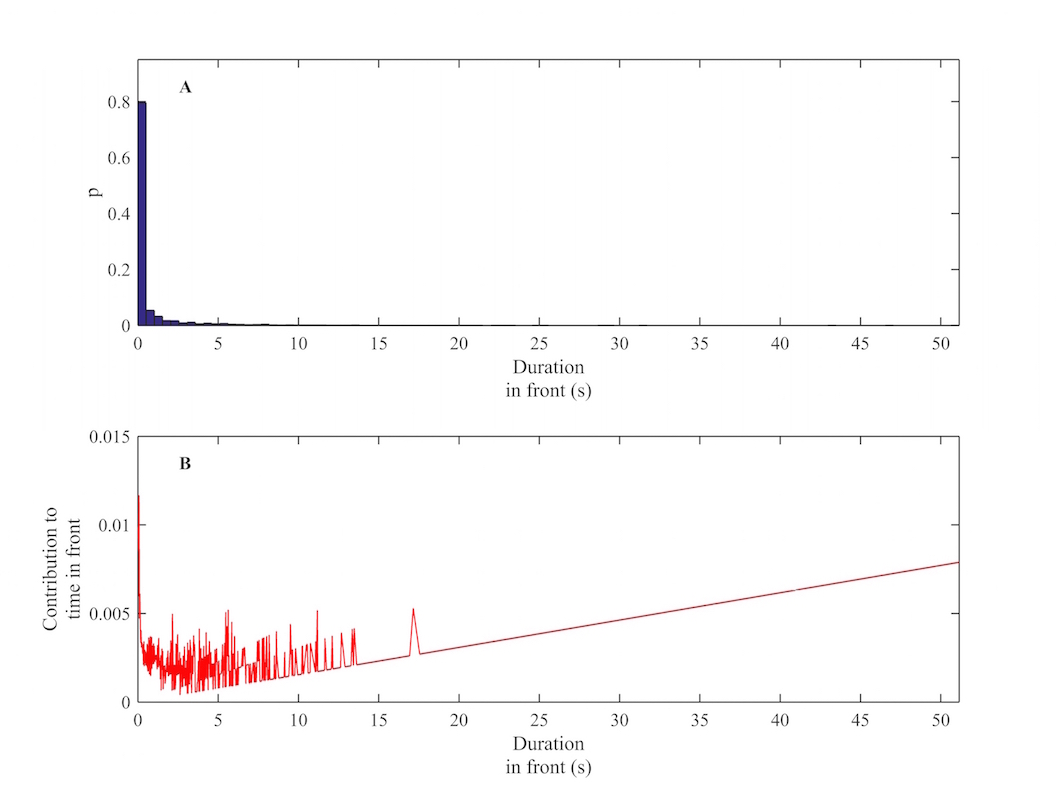}
	\end{center}
	\caption{(A) The relative frequency of unbroken durations (\emph{segments}) spent at the front of the pair by either a front fish or back fish. Note there are many instances of individuals only occupying the front position for short periods of time.  However, as shown in B, these short switches contribute negligible amounts of data to our analysis. (B) shows the contribution that segments of different length ($x$-axis) make towards the data set.} \label{fig:individualdurationsinfront}
\end{figure}

During our analysis we noted that many instances of occupying the frontmost position only lasted for short durations. When the front position was in contest, the frontmost fish relative to the group centroid would often swap multiple times. Figure \ref{fig:individualdurationsinfront} A illustrates the relative frequency that unbroken durations spent at the front of the pair by either a front fish or back fish were observed. The histogram is dominated by short duration instances of occupying the front. However, the large number of short duration instances of front position occupancy only contributed a small amount to the total duration of data that we analysed. Figure \ref{fig:individualdurationsinfront} B illustrates the proportion of the total data analysed that were made up of instances when a fish occupied the front position for a given duration. The shortest duration instances (of 0.025 seconds = 1 frame) only made up 0.99 \% of the total data analysed, durations of front occupancy of 0.1 seconds (4 frames ) or less made up approximately 3.48 \% of the data analysed and durations of 1.0 seconds (40 frames) or less made up approximately 13.63 \% of the data analysed.   

\begin{table}[!h]
	\begin{center}
	\caption{The total number of frames where a pair of fish was closely grouped (within 100 mm of each other), and the proportion of frames spent at the front of the pair by each fish. For these groups fish 1 had an additional 5 minutes to familiarise itself with the tank before fish 2 was released. Bold type indicates that a given fish spent the largest proportion of frames at the front of the pair.}\label{tab:proportions1}
	\begin{tabular}{llll}	
	\hline\noalign{\smallskip}
	Group & No. frames & Prop. fish 1 & Prop. fish 2 \\
	& closely grouped & in front & in front \\
	\noalign{\smallskip}
	\hline
	1 & 11655 & {\bf 0.5887} & 0.4113  \\
	2 & 11690 & {\bf 0.5377} & 0.4623  \\
	3 & 9844 & {\bf 0.6790} & 0.3210  \\
	4 & 11645 & {\bf 0.6670} & 0.3330  \\
	5 & 12671 & 0.4686 & {\bf 0.5314}  \\
	6 & 3704 & 0.2125 & {\bf 0.7875}  \\
	7 & 1340 & {\bf 0.5657} & 0.4343  \\
	8 &	9917 & {\bf 0.5301} & 0.4699  \\
	9 &	10120 & 0.4941 & {\bf 0.5059}  \\
	10 & 10402 & {\bf 0.8101} & 0.1899  \\
	11 & 5019 & 0.4082 & {\bf 0.5918}  \\
	12 & 6565 & {\bf 0.7555} & 0.2445  \\
	13 & 1147 & {\bf 0.5004} & 0.4996  \\
	14 & 961 & 0.3018 & {\bf 0.6982}  \\
	15 & 7694 & 0.4082 & {\bf 0.5918}  \\
	16 & 3537 & 0.4515 & {\bf 0.5485}  \\
	17 & 10110 & 0.3910 & {\bf 0.6090}  \\
	18 & 4569 & 0.4730 & {\bf 0.5270}  \\
	19 & 595 & 0.4723 & {\bf 0.5277}  \\
	20 & 9093 & 0.4576 & {\bf 0.5424}  \\
	\hline
	\end{tabular}
	\end{center}
\end{table}

\begin{table}[!h]
	\begin{center}
	\caption{The total number of frames where a pair of fish was closely grouped (within 100 mm of each other), and the proportion of frames spent at the front of the pair by each fish. For these groups both fish were released simultaneously. Bold type indicates that a given fish spent the largest proportion of frames at the front of the pair.}\label{tab:proportions2}
	\begin{tabular}{llll}	
	\hline\noalign{\smallskip}
	Group & No. frames & Prop. fish 1 & Prop. fish 2 \\
	& closely grouped & in front & in front \\
	\noalign{\smallskip}
	\hline
	21 & 8133 & {\bf 0.6957} & 0.3043 \\
	22 & 6870 & 0.0357 & {\bf 0.9643} \\
	23 & 8748 & {\bf 0.8994} & 0.1006 \\
	24 & 9820 & {\bf 0.7011} & 0.2989 \\
	25 & 10714 & 0.2976 & {\bf 0.7024} \\
	26 & 4762 & 0.4794 & {\bf 0.5206} \\
	27 & 3811 & 0.3836 & {\bf 0.6164} \\
	28 & 5761 & {\bf 0.6804} & 0.3196 \\
	29 & 5954 & 0.2538 & {\bf 0.7462} \\
	30 & 9157 & 0.3374 & {\bf 0.6626} \\
	31 & 6301 & 0.0689 & {\bf 0.9311} \\
	32 & 7871 & {\bf 0.9625} & 0.0375 \\
	33 & 5202 & 0.3754 & {\bf 0.6246} \\
	34 & 3912 & {\bf 0.5378} & 0.4622 \\
	35 & 3091 & {\bf 0.7383} & 0.2617 \\
	36 & 242 & 0.4793 & {\bf 0.5207} \\
	37 & 2962 & {\bf 0.7269} & 0.2731 \\
	38 & 2887 & 0.4288 & {\bf 0.5712} \\ 
	39 & 7434 & 0.1987 & {\bf 0.8013} \\
	40 & 6872 & {\bf 0.8615} & 0.1385  \\
	\hline
	\end{tabular}
	\end{center}
\end{table}

\clearpage

\subsection*{Time lag associated with maximum directional correlation}\label{supp:dcdtables}
\begin{figure}[!h]
	\begin{center}
	\includegraphics[width=\textwidth]{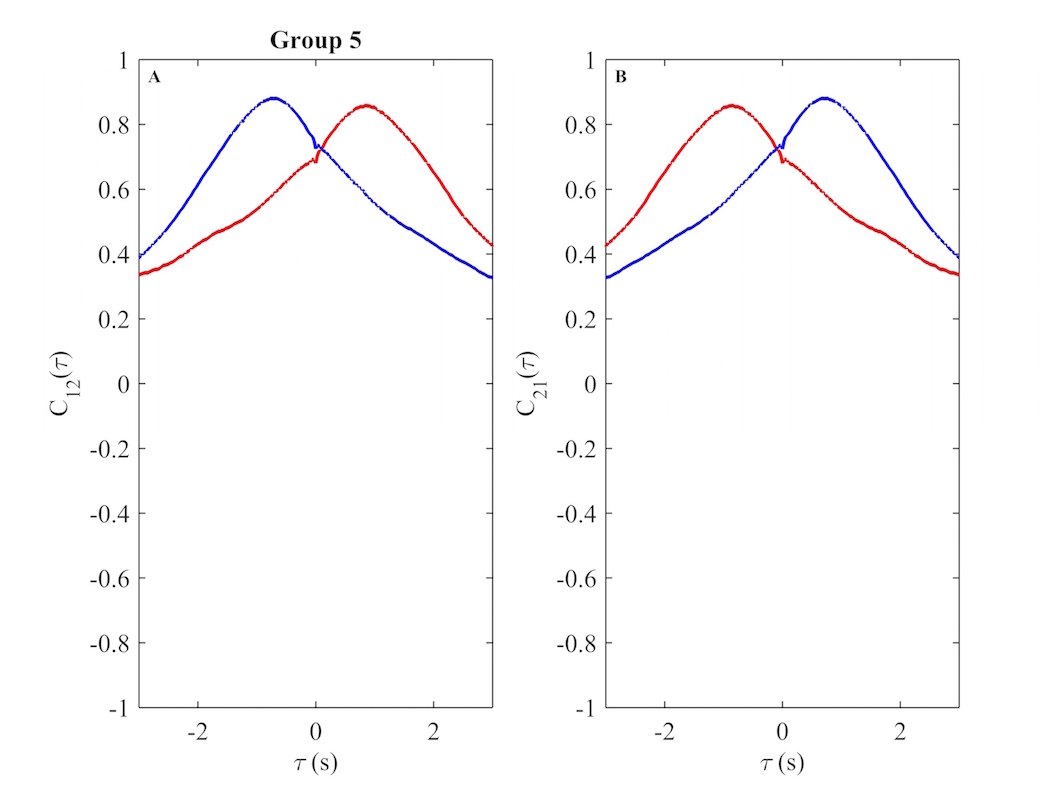}
	\end{center}
	\caption{Directional correlation, $C_{ij}(\tau)$, for group 5 when fish 1 was in front (red lines) and when fish 2 was in front (blue lines). The horizontal scale in both plots is time lag, $\tau$, measured in seconds. (A) illustrates the mean correlation in direction of motion between fish 1 at time $t$ with the direction of motion of fish 2 at time $t+\tau$. (B) illustrates the mean correlation in direction of motion between fish 2 at time $t$ with the direction of motion of fish 1 at time $t+\tau$.}\label{fig:dcd_example}
\end{figure}

Tables \ref{tab:dcd1} and \ref{tab:dcd2} detail the time lag associated with maximum directional correlation and the corresponding maximum directional correlation for all 40 groups of fish. Figure \ref{fig:dcd_example} contains an example of directional correlation delay plots for periods spent at the front and rear of a given pair of fish (group 5). 

\begin{table}[!h]
	\begin{center}
	\caption{Time lag associated with maximum directional correlation, $\tau^{*}_{ij}$, and maximum directional correlation, $C_{ij}\left(\tau^{*}_{ij}\right)$, for when either fish 1 or fish 2 occupied the front-most position of the group (groups 1 to 20).}\label{tab:dcd1}
	\begin{tabular}{lllllll}
	\hline\noalign{\smallskip}
	 & \multicolumn{4}{c}{$\tau^{*}_{ij}$} & \multicolumn{2}{c}{$C_{ij}\left(\tau^{*}_{ij}\right)$}  \\
	 \noalign{\smallskip}
	Group & Fish 1 & Fish 1 & Fish 2 & Fish 2 & Fish 1 & Fish 1 \\
	 & in front & behind & in front & behind & in front & behind \\
	 & & & & & (Fish 2 & (Fish 2 \\
	 & & & & & behind) & in front) \\
	\noalign{\smallskip}
	\hline
	1 & 0.700 & -0.550 & 0.550 & -0.700 & 0.9335 & 0.9101 \\
	2 & 0.875 & -0.875 & 0.875 & -0.875 & 0.8828 & 0.8996 \\
	3 & 0.750 & -0.875 & 0.875 & -0.750 & 0.9162 & 0.8894 \\
	4 & 0.950 & -0.650 & 0.650 & -0.950 & 0.9242 & 0.8926 \\
	5 & 0.850 & -0.675 & 0.675 & -0.850 & 0.8588 & 0.8807 \\
	6 & 1.475 & -1.100 & 1.100 & -1.475 & 0.9177 & 0.7403 \\
	7 & -3.000 & -2.475 & 2.475 & 3.000 & 0.6368 & 0.9853 \\
	8 & 0.700 & -0.675 & 0.675 & -0.700 & 0.9009 & 0.8871 \\
	9 & 0.825 & -0.900 & 0.900 & -0.825 & 0.8298 & 0.8827 \\
	10 & 0.625 & -0.875 & 0.875 & -0.625 & 0.9418 & 0.8225 \\
	11 & 1.125 & -0.400 & 0.400 & -1.125 & 0.8804 & 0.8956 \\
	12 & 0.575 & -0.875 & 0.875 & -0.575 & 0.8783 & 0.7246 \\
	13 & 1.575 & -1.575 & 1.575 & -1.575 & 0.8755 & 0.7979 \\
	14 & -2.400 & -1.150 & 1.150 & 2.400 & 0.9734 & 0.7829 \\
	15 & 0.675 & -0.675 & 0.675 & -0.675 & 0.9246 & 0.8983 \\
	16 & 1.050 & -0.675 & 0.675 & -1.050 & 0.8940 & 0.8306 \\
	17 & 0.700 & -0.700 & 0.700 & -0.700 & 0.8587 & 0.9127 \\
	18 & 1.200 & -1.150 & 1.150 & -1.200 & 0.7852 & 0.7880 \\
	19 & 1.000 & -1.050 & 1.050 & -1.000 & 0.4320 & 0.7499 \\
	20 & 0.925 & -0.800 & 0.800 & -0.925 & 0.8497 & 0.8718 \\
	\hline
	\end{tabular}
	\end{center}
\end{table}

\begin{table}[!h]
	\begin{center}
	\caption{Time lag associated with maximum directional correlation, $\tau^{*}_{ij}$, and maximum directional correlation, $C_{ij}\left(\tau^{*}_{ij}\right)$, for groups 21 to 40.}\label{tab:dcd2}
	\begin{tabular}{lllllll}
	\hline\noalign{\smallskip}
	 & \multicolumn{4}{c}{$\tau^{*}_{ij}$} & \multicolumn{2}{c}{$C_{ij}\left(\tau^{*}_{ij}\right)$}  \\
	 \noalign{\smallskip}
	Group & Fish 1 & Fish 1 & Fish 2 & Fish 2 & Fish 1 & Fish 1 \\
	 & in front & behind & in front & behind & in front & behind  \\
	 & & & & & (Fish 2 & (Fish 2 \\
	 & & & & & behind) & in front) \\
	\noalign{\smallskip}
	\hline
	21 & 1.125 & -1.300 & 1.300 & -1.125 & 0.8492 & 0.7333 \\
	22 & 2.400 & -0.700 & 0.700 & -2.400 & 0.8634 & 0.9520 \\
	23 & 0.600 & -0.850 & 0.850 & -0.600 & 0.9613 & 0.8427 \\
	24 & 0.700 & -0.700 & 0.700 & -0.700 & 0.9223 & 0.8983 \\
	25 & 1.875 & -1.125 & 1.125 & -1.875 & 0.7638 & 0.8098 \\
	26 & 0.900 & -1.000 & 1.000 & -0.900 & 0.9382 & 0.9142 \\
	27 & 0.975 & -0.950 & 0.950 & -0.975 & 0.8836 & 0.8850 \\
	28 & 0.825 & -1.075 & 1.075 & -0.825 & 0.8932 & 0.9026 \\
	29 & 0.875 & -0.675 & 0.675 & -0.875 & 0.9183 & 0.9445 \\
	30 & 0.675 & -0.575 & 0.575 & -0.675 & 0.9197 & 0.9194 \\
	31 & -2.650 & -0.650 & 0.650 & 2.650 & 0.6307 & 0.8182 \\
	32 & 0.650 & 0.875 & -0.875 & -0.650 & 0.9093 & 0.8933 \\
	33 &-0.900 & -1.300 & 1.300 &  0.900 & 0.6792 & 0.7006 \\
	34 & 1.000 & -1.100 & 1.100 & -1.000 & 0.8085 & 0.7562 \\
	35 & 0.800 & 1.725 & -1.725 & -0.800 & 0.8166 & 0.8767 \\
	36 &-1.325 & -1.050 & 1.050 &  1.325 & 0.0358 & -0.4376 \\
	37 & 0.700 & -1.150 & 1.150 & -0.700 & 0.9152 & 0.8622 \\
	38 & 0.675 & -0.725 & 0.725 & -0.675 & 0.9015 & 0.9331 \\
	39 & 0.900 & -0.775 & 0.775 & -0.900 & 0.8406 & 0.9388 \\
	40 & 0.800 & -0.825 & 0.825 & -0.800 & 0.9386 & 0.8371 \\
	\hline
	\end{tabular}
	\end{center}
\end{table}

\clearpage

\subsection*{Comparison of basic movement statistics}\label{supp:basicstatstable}
Table \ref{tab:basictallies1} summarises the results of statistical tests to determine if there were any differences in summary statistics for properties of movement or body length between individuals that occupied the frontmost position of the pair when closely grouped (within 100 mm of each other). Test results were sorted in ascending order of $p$-value, and significance levels for each test were adjusted according to a Holm-Bonferroni correction \citealt{Holm1979} (see the fifth column of table \ref{tab:basictallies1}). In the absence of a Holm-Bonferroni correction, fish that occupied the frontmost position for the greatest duration (compared to their partner) differed from their partner in mean change in speed over time (median difference (front fish minus back fish) in mean change in speed over time = 6.5340 mm/s$^2$), inter-quartile range of speed (mean difference in IQR of speed = 5.0105 mm/s), body length (mean difference in body length = 1.0608 mm), median turning speed (median difference in median turning speed = -6.2436$^\circ$/s) and standard deviation in speed (median difference in standard deviation of speed = 1.5169 mm/s). With a Holm-Bonferroni correction active, only differences in mean change in speed over time remained significant.   

\begin{table}[!h]
	\begin{center}
	\tiny
	\caption{Results of paired statistical tests applied to summary statistics of locomotive properties and body lengths of fish that occupied the front or back of their pair for the greatest proportion of frames when closely grouped ($\leq 100$ mm from each other). (Locomotive properties examined were speed ($s_{i}(t)$), change in speed over time ($\frac{\Delta s_{i}}{\Delta t}$), magnitude of acceleration ($a_{i}(t)$) and turning speed ($\alpha_{i}(t)$).) Differences between each summary statistic for each pair of fish were first determined. The distribution of these differences were tested for departures from normality using a Shapiro-Wilk test. If the differences were likely to have been drawn from a normal distribution, then data was further tested using a paired $t$-test (to determine if the mean of the differences differed from zero). If the differences were not normally distributed, then data was further tested according to a Wilcoxon signed rank test as applied by MATLAB's intrinsic \emph{signrank} function (to determined if the median of the differences departed from zero). Test results are sorted in ascending $p$-value, with appropriate significance levels $\alpha_{\textrm{sig}}$ adjusted according to a Holm-Bonferroni correction listed in the fourth column. All $t$-tests had 39 degrees of freedom.}\label{tab:basictallies1} 
	\begin{tabular}{lllll}
	\hline\noalign{\smallskip}
	Summary & Method of & Test statistics & $p$ & $\alpha_{\textrm{sig}}$ \\
	statistic & comparison &  &  & \\
	\noalign{\smallskip}
	\hline
	Mean $\frac{\Delta s_{i}}{\Delta t}(t)$ & signed rank test & $W=716$, $z=4.1130$ & $3.9049 \times 10^{-5}$ & 0.024 \\
	IQR $s_{i}(t)$ & $t$-test & $\tau = 3.0744$ & 0.0038 & 0.0025 \\
	Body length & $t$-test & $\tau = 2.9633$ & 0.0052 & 0.0026 \\
	Median $\alpha_{i}(t)$ & signed rank test & $W=237$, $z=-2.3253$ & 0.0201 & 0.0028 \\
	Std $s_{i}(t)$ & signed rank test & $W=556$, $z=1.9624$ & 0.0497 & 0.0029 \\
	Median $s_{i}(t)$ & signed rank test & $W=532$, $z=1.6398$ & 0.1010 & 0.0031 \\
	Mean $s_{i}(t)$ & signed rank test & $W=518$, $z=1.4517$ & 0.1466 & 0.0033 \\
	IQR $\frac{\Delta s_{i}}{\Delta t}(t)$ & signed rank test & $W=303$, $z=-1.4382$ & 0.1504 & 0.0036 \\
	Median $a_{i}(t)$ & $t$-test & $\tau=-1.3317$ & 0.1907 & 0.0038 \\
	IQR $\alpha_{i}(t)$ & signed rank test & $W=325$, $z=-1.1425$ & 0.2532 & 0.0042 \\
	Median $\frac{\Delta s_{i}}{\Delta t}(t)$ & signed rank test & $W=482$, $z=0.9678$ & 0.3332 & 0.0045 \\
	Maximum $a_{i}(t)$ & signed rank test & $W=469$, $z=0.7930$ & 0.4278 & 0.0050 \\
	IQR $a_{i}(t)$ & signed rank test & $W=366$, $z=-0.5914$ & 0.5542 & 0.0056 \\
	Mean $a_{i}(t)$ & $t$-test & $\tau=-0.5151$ & 0.6094 & 0.0063 \\
	Mean $\alpha_{i}(t)$ & signed rank test & $W=382$, $z=-0.3764$ & 0.7067 & 0.0071 \\
	Std $\alpha_{i}(t)$ & $t$-test & $\tau=0.3718$ & 0.7120 & 0.0083 \\
	Std $\frac{\Delta s_{i}}{\Delta t}(t)$ & $t$-test & $\tau=-0.2848$ & 0.7773 & 0.0100 \\
	Maximum $\alpha_{i}(t)$ & signed rank test & $W=211$, $z=-0.1406$ & 0.8882 & 0.0125 \\
	Maximum $s_{i}(t)$ & signed rank test & $W=416$, $z=0.0806$ & 0.9357 & 0.0167 \\	
	Maximum $\frac{\Delta s_{i}}{\Delta t}(t)$ & signed rank test & $W=405$, $z=-0.0672$ & 0.9464 & 0.0250 \\	
	Std $a_{i}(t)$ & signed rank test & $W=407$, $z=-0.0403$ & 0.9678 & 0.0500 \\
	\hline
	\end{tabular}
	\end{center}
\end{table}

\clearpage

\subsection*{Distributions of movement parameters and body lengths}\label{supp:distributions}

\begin{figure}[!h]
	\begin{center}
	\includegraphics[width=\textwidth]{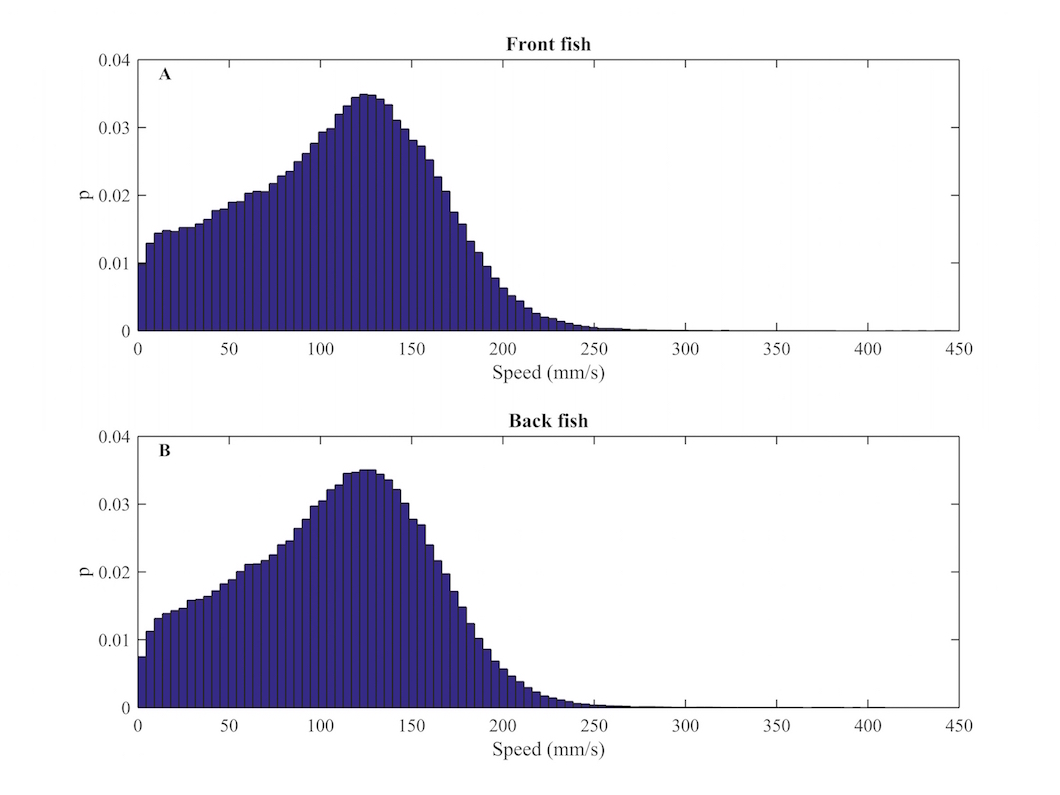}
	\end{center}
	\caption{The distribution of observed speeds (in mm/s) for fish that spent the greatest proportion of their time at the front (A) or back (B) of each pair when the fish were $\leq 100$ mm apart (pooled from all 40 trials).}\label{fig:speed_hist}
\end{figure}

\begin{figure}[!h]
	\begin{center}
	\includegraphics[width=\textwidth]{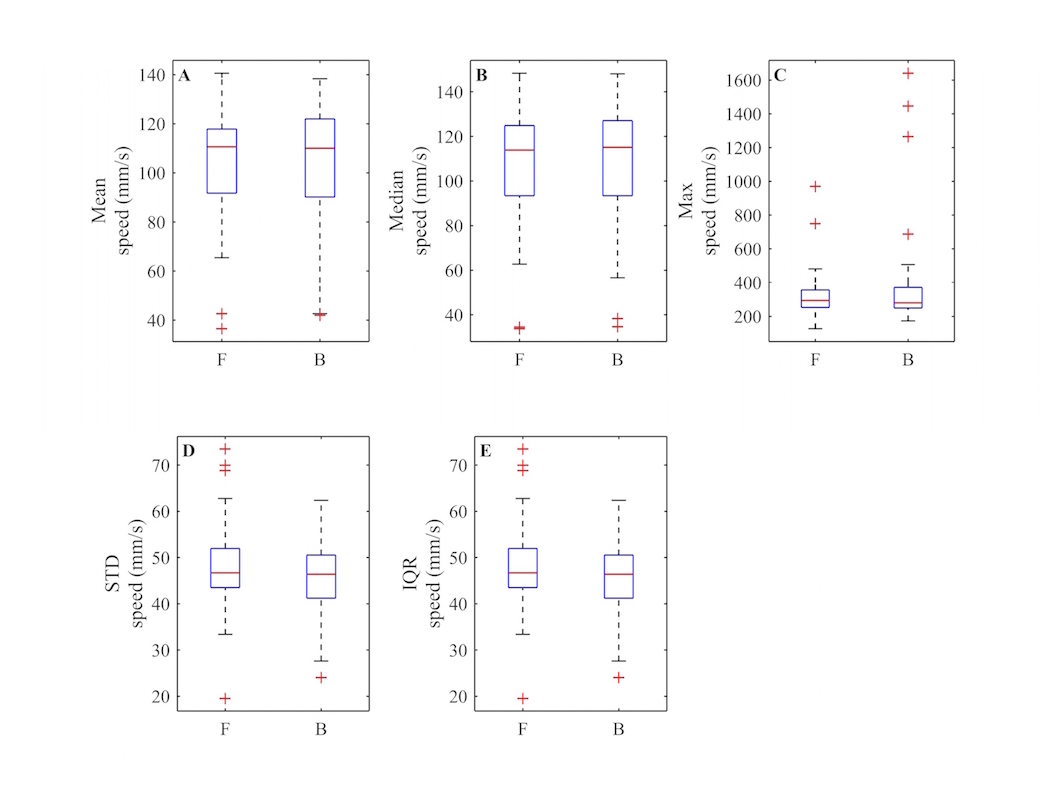}
	\end{center}
	\caption{Boxplots of the mean (A), median (B), maximum (C), standard deviation (D) and inter-quartile range (E) of speed (in mm/s) for fish that occupied the front (F) or back (B) of a pair for the greatest duration when the fish were $\leq 100$ mm apart.}\label{fig:speed_boxplot}
\end{figure}

\begin{figure}[!h]
	\begin{center}
	\includegraphics[width=\textwidth]{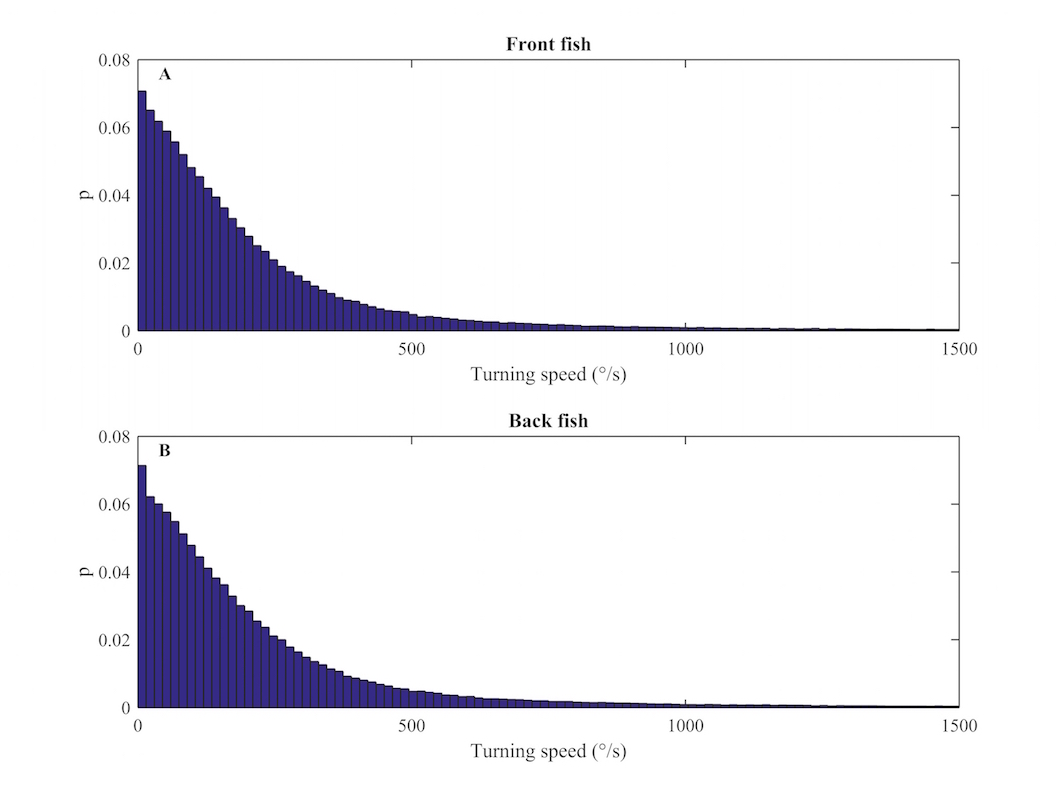}
	\end{center}
	\caption{The distribution of observed turning speeds (in $^{\circ}$/s) for fish that spent the greatest proportion of their time at the front (A) or back (B) of each pair when the fish were $\leq 100$ mm apart (pooled from all 40 trials).}\label{fig:turningspeed_hist}
\end{figure}

\begin{figure}[!h]
	\begin{center}
	\includegraphics[width=\textwidth]{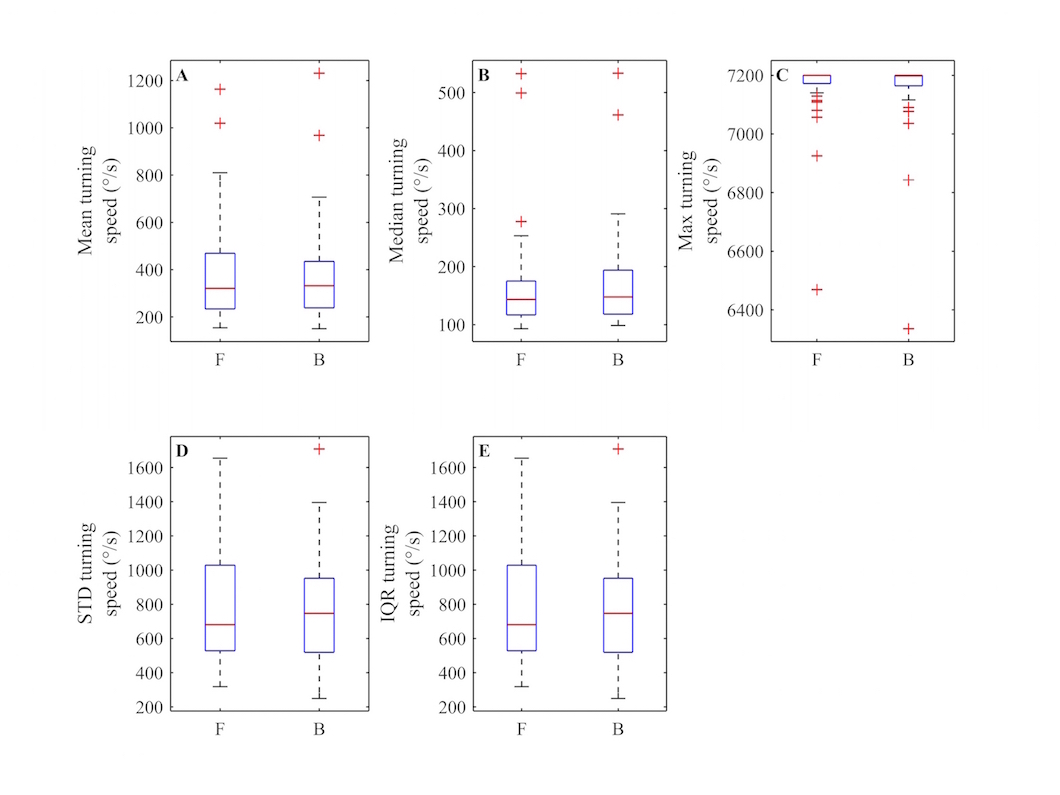}
	\end{center}
	\caption{Boxplots of the mean (A), median (B), maximum (C), standard deviation (D) and inter-quartile range (E) of turning speed (in $^{\circ}$/s) for fish that occupied the front (F) or back (B) of a pair for the greatest duration when the fish were $\leq 100$ mm apart.}\label{fig:turningspeed_boxplot}
\end{figure}

\begin{figure}[!h]
	\begin{center}
	\includegraphics[width=\textwidth]{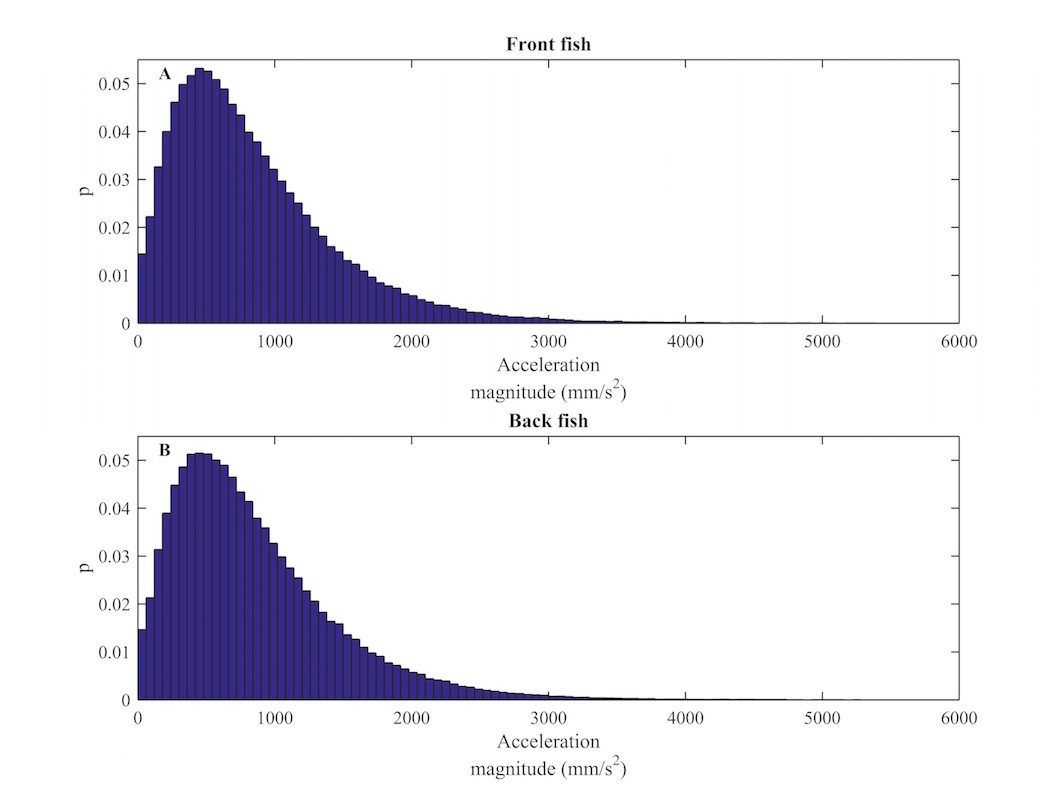}
	\end{center}
	\caption{The distribution of observed magnitudes of acceleration (in mm/s$^2$) for fish that spent the greatest proportion of their time at the front (A) or back (B) of each pair when the fish were $\leq 100$ mm apart (pooled from all 40 trials).}\label{fig:accelerationmagnitude_hist}
\end{figure}

\begin{figure}[!h]
	\begin{center}
	\includegraphics[width=\textwidth]{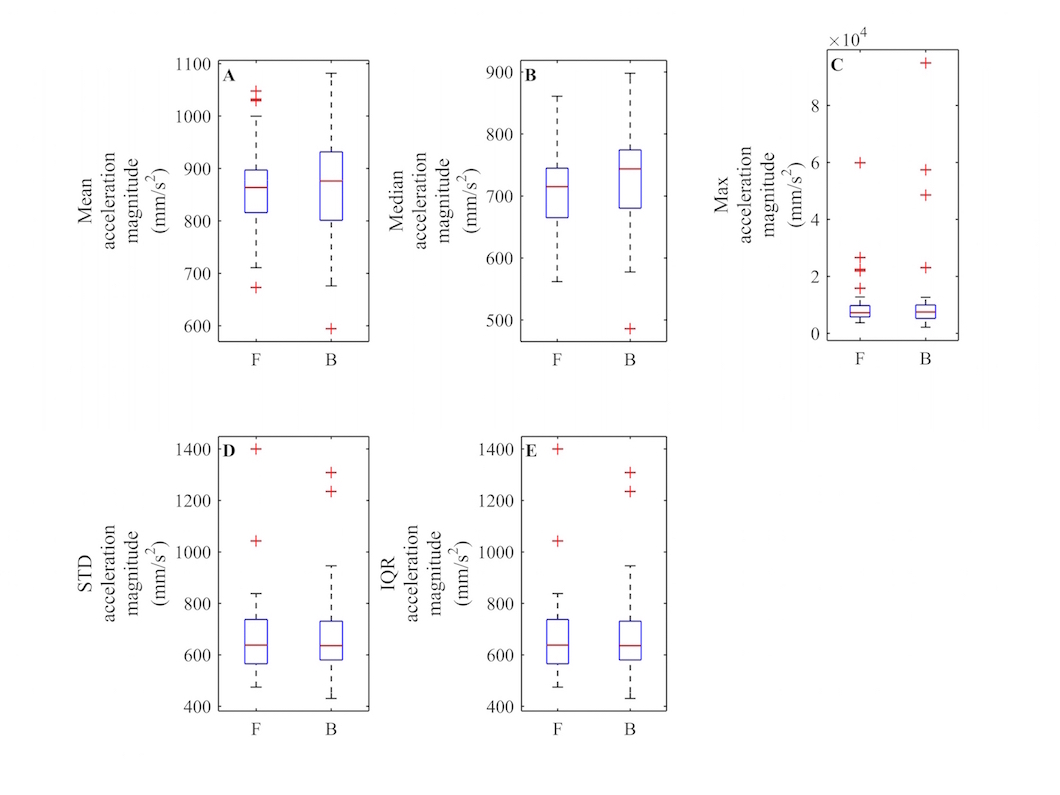}
	\end{center}
	\caption{Boxplots of the mean (A), median (B), maximum (C), standard deviation (D) and inter-quartile range (E) of magnitudes of acceleration (in mm/s$^2$) for fish that occupied the front (F) or back (B) of a pair for the greatest duration when the fish were $\leq 100$ mm apart.}\label{fig:accelerationmagnitude_boxplot}
\end{figure}

\begin{figure}[!h]
	\begin{center}
	\includegraphics[width=\textwidth]{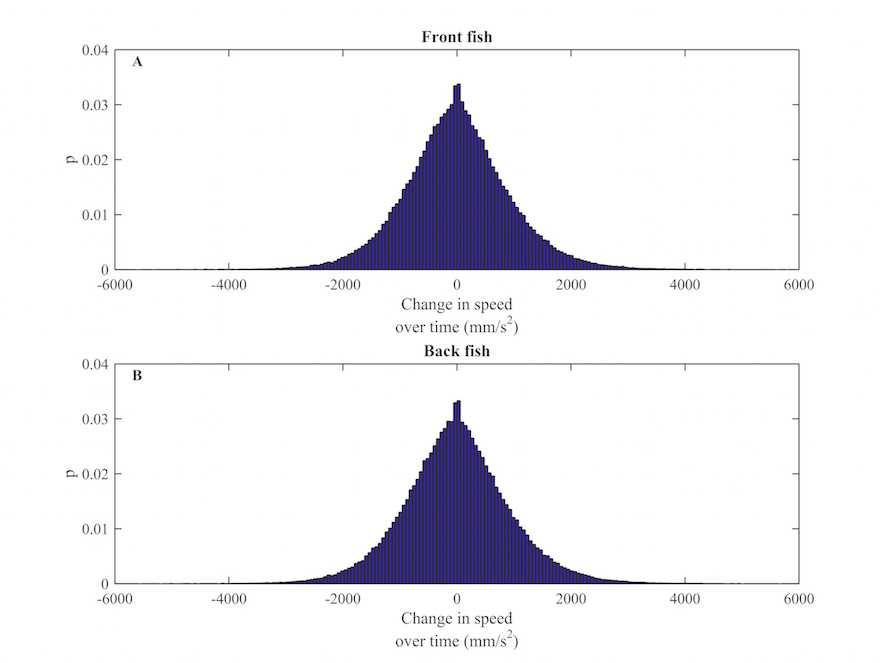}
	\end{center}
	\caption{The distribution of observed changes in speed over time (in mm/s$^2$) for fish that spent the greatest proportion of their time at the front (A) or back (B) of each pair when the fish were $\leq 100$ mm apart (pooled from all 40 trials).}\label{fig:changeinspeed_hist}
\end{figure}

\begin{figure}[!h]
	\begin{center}
	\includegraphics[width=\textwidth]{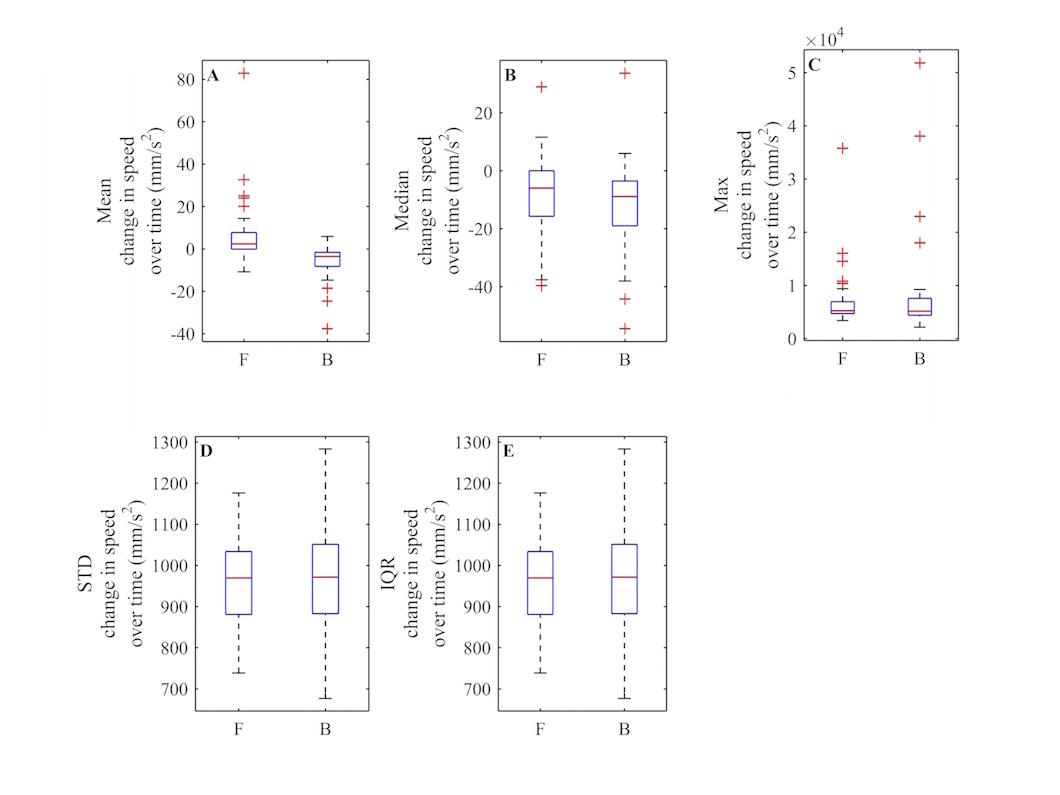}
	\end{center}
	\caption{Boxplots of the mean (A), median (B), maximum (C), standard deviation (D) and inter-quartile range (E) of changes in speed over time (in mm/s$^2$) for fish that occupied the front (F) or back (B) of a pair for the greatest duration when the fish were $\leq 100$ mm apart.}\label{fig:changeinspeed_boxplot}
\end{figure}

\begin{figure}[!h]
	\begin{center}
	\includegraphics[width=\textwidth]{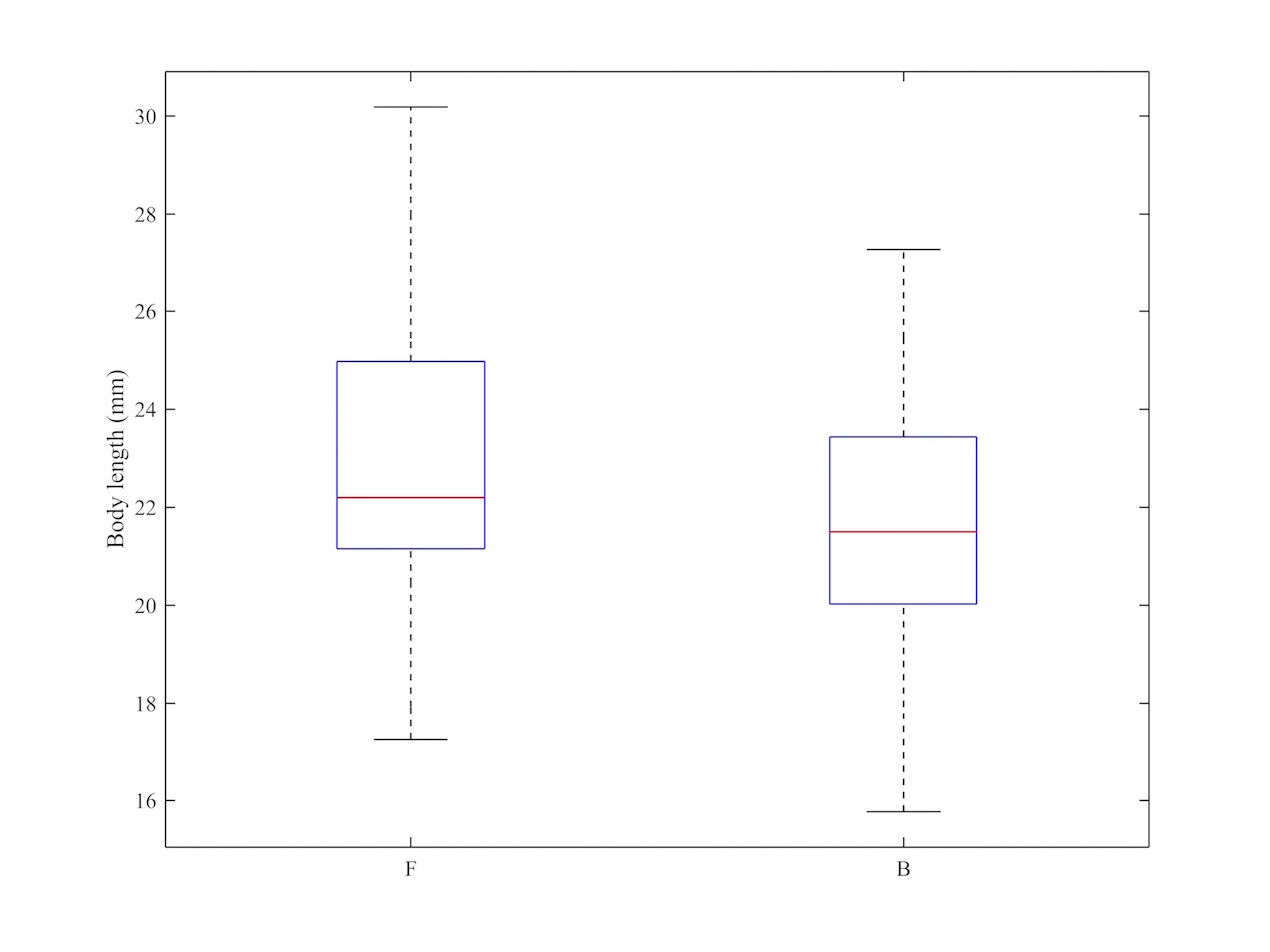}
	\end{center}
	\caption{Boxplots of body length (in mm) for fish that occupied the front (F) or back (B) of a pair for the greatest duration when the fish were $\leq 100$ mm apart.}\label{fig:bodylengths_boxplot}
\end{figure}

\begin{figure}[!h]
	\begin{center}
	\includegraphics[width=\textwidth]{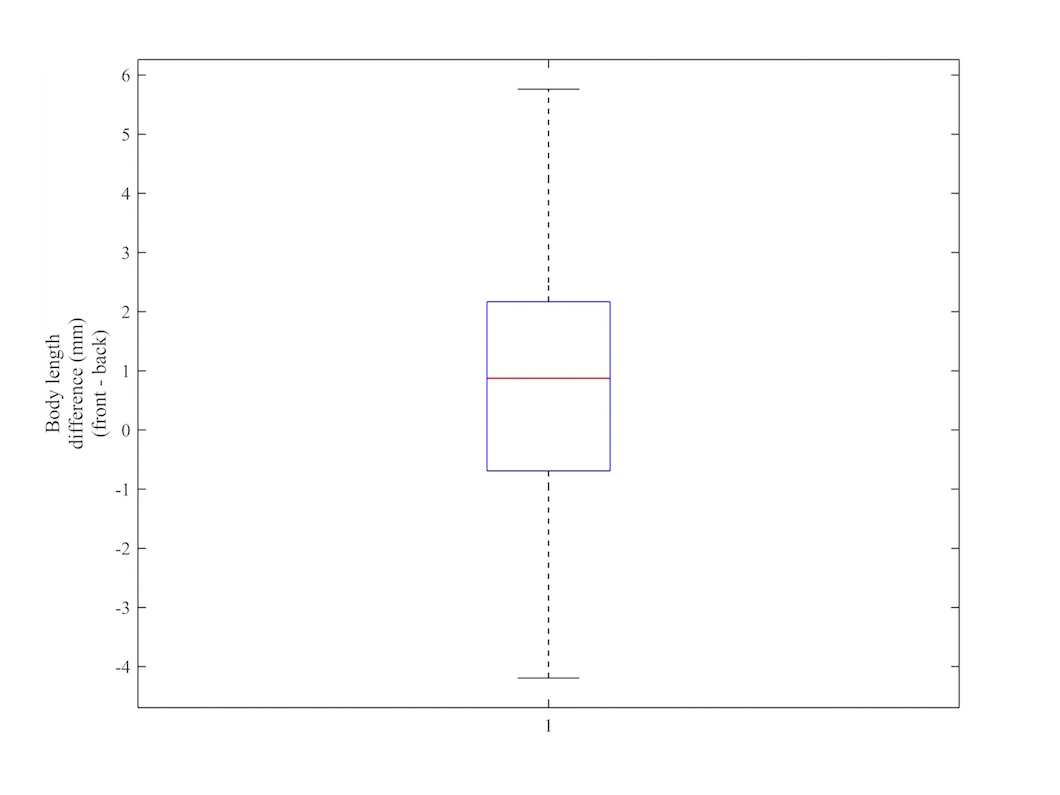}
	\end{center}
	\caption{Boxplot of the difference in body length (in mm) between fish that occupied the front (F) or back (B) of a pair for the greatest duration when the fish were $\leq 100$ mm apart. The median body length difference was 0.8748 mm, and the mean body length difference was 1.0608 mm (which differed from 0 according to a paired $t$-test, $p = 0.0052$).}\label{fig:bodylengthdiffs_boxplot}
\end{figure}

\clearpage

\subsection*{Proportion of encounters, speed, relative direction of motion and rules of motion as a function of relative neighbour location}\label{supp:prop_speed_align}

\begin{figure}[!h]
	\begin{center}
	\includegraphics[width=\textwidth]{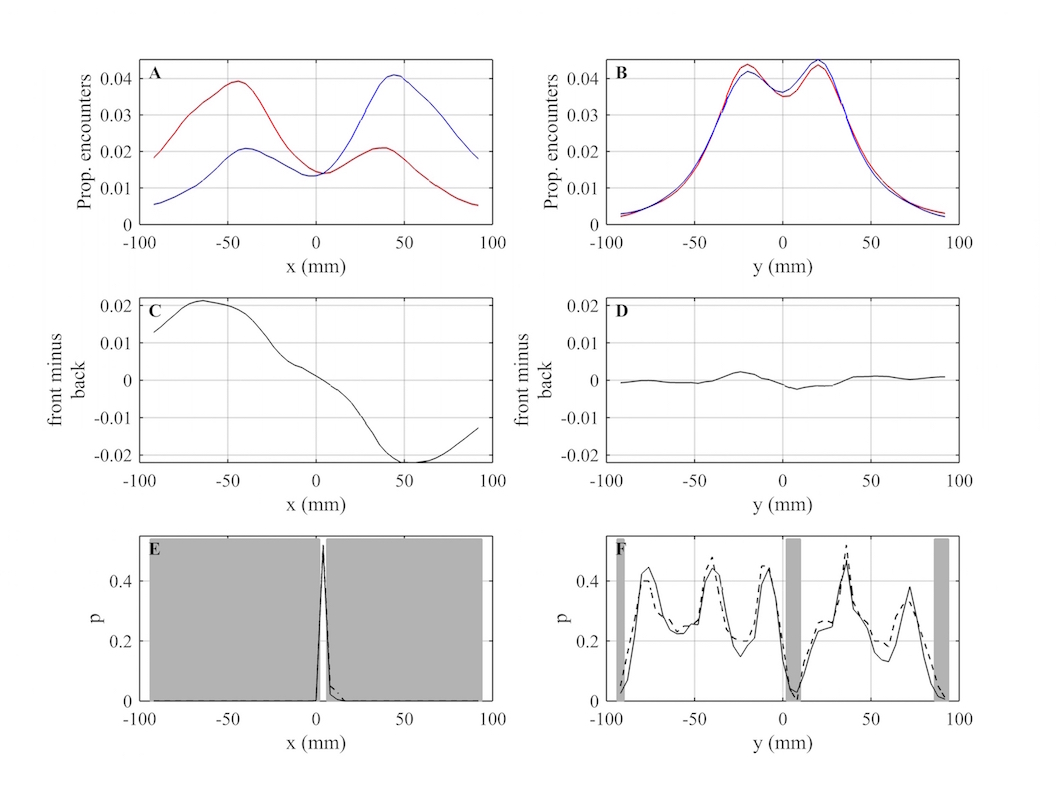}
	\end{center}
	\caption{Proportion of total encounters with partner fish projected onto the $x$-axis (A) and the $y$-axis (B). $x$ or $y$ coordinates on the horizontal axis of the graphs represent the $x$ or $y$ coordinates of partner fish relative to the location and direction of motion of a given focal fish type. The curves show the proportion of encounters where a front fish's (red curve) or back fish's (blue curve) neighbour was observed. The middle panels illustrate the front fish minus the back fish proportion of encounters projected onto the $x$ (C) and $y$-axes (D). (E) and (F) illustrate the estimated probability of a difference at least the same magnitude and sign as that observed occurring if fish were randomly allocated to the set of front fish or back fish. Probabilities estimated from a set of 100 randomisation processes are plotted as dashed lines and probabilities derived from 1000 randomisations are plotted as solid lines. Grey regions indicate where estimated probabilities derived from 1000 randomisation processes are less than 0.05.}\label{fig:propencounters_projections}
\end{figure}

\begin{figure}[!h]
	\begin{center}
	\includegraphics[width=\textwidth]{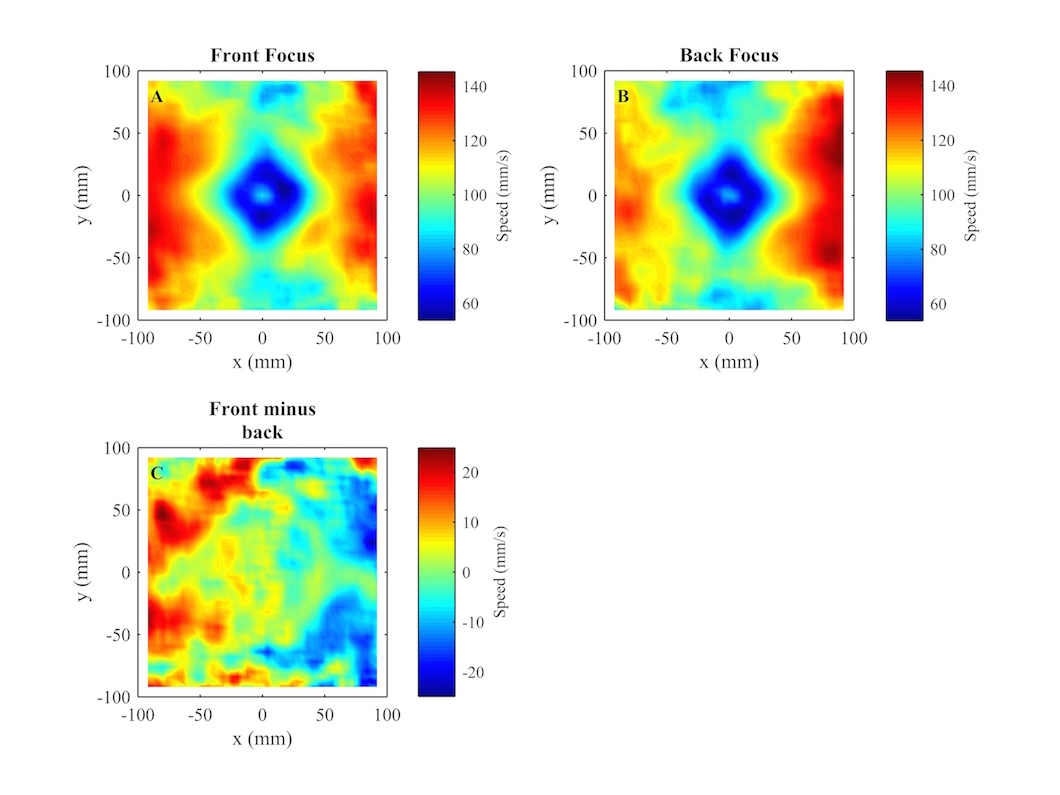}
	\end{center}
	\caption{(A) and (B) show heat maps of mean speed (mm/s) as a function of the location of partner fish relative to the direction of motion and position of a focal fish type (front fish or back fish). Focal fish are located at the origin of each plot, moving in the direction of the positive $x$-axis. (C) contains the difference obtained from the front fish focused map minus the back fish focused map. Back fish travel slower than front fish when their partner is behind them.}\label{fig:speed_fxy}
\end{figure}

\begin{figure}[!h]
	\begin{center}
	\includegraphics[width=\textwidth]{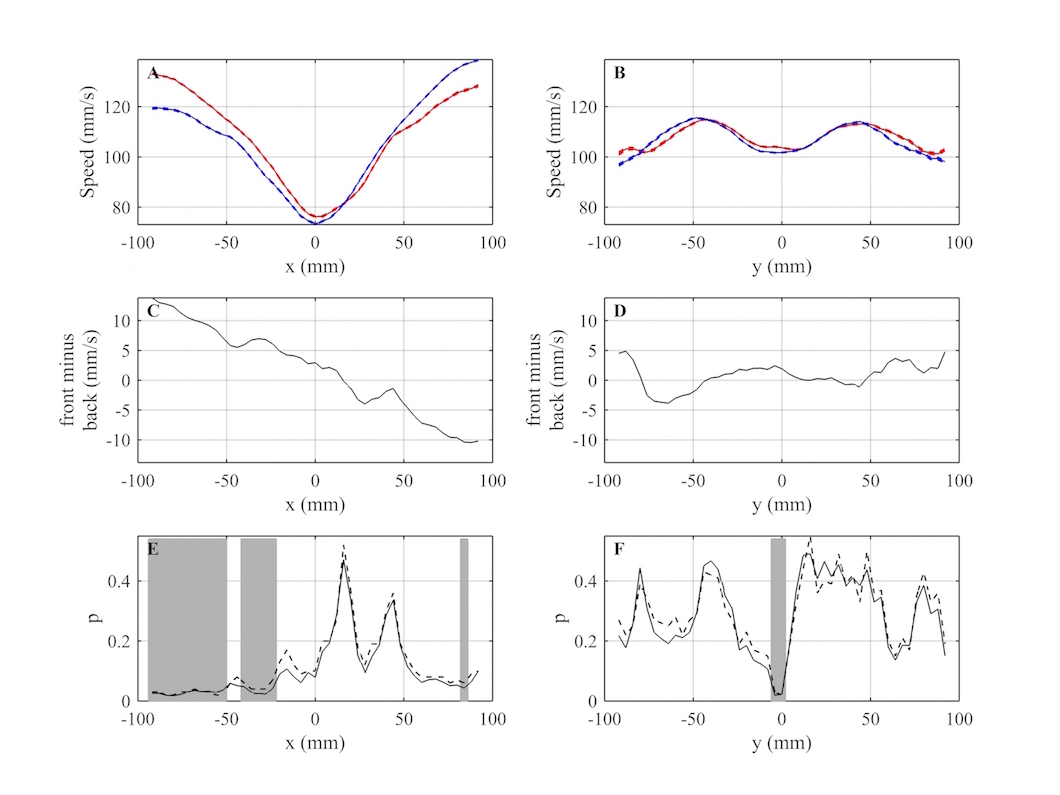}
	\end{center}
	\caption{Mean speed of focal fish as a function of their partner's $x$- (A) or $y$-coordinate (B). Curves representing the mean speed of front fish are plotted in red; curves representing the mean speed of back fish are plotted in blue. Dashed lines are plotted one standard error above and below mean speed curves. (C) and (D) illustrate the difference in front fish and back fish curves, and (E) and (F) illustrate the estimated probability of a difference at least the same magnitude and sign as that observed occurring. Probabilities estimated from a set of 100 randomisation processes are plotted as dashed lines and probabilities derived from 1000 randomisations are plotted as solid lines. Grey regions indicate where estimated probabilities derived from 1000 randomisation processes are less than 0.05. Back fish are slower than front fish when their partner is behind them.}\label{fig:speed_projections}
\end{figure}

\begin{figure}[!h]
	\begin{center}
	\includegraphics[width=\textwidth]{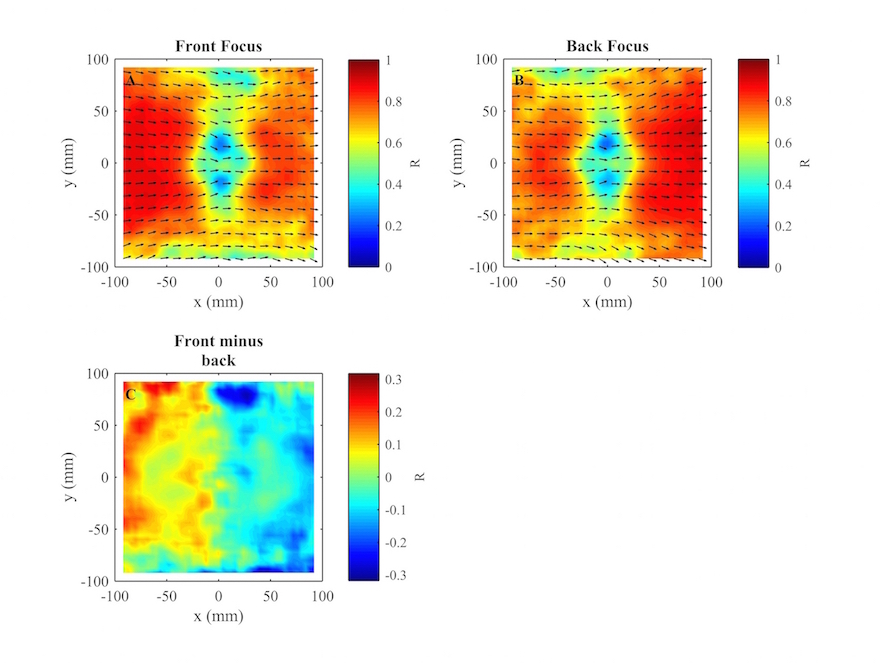}
	\end{center}
	\caption{(A) and (B) show heat maps of $R$ (given by equation (\ref{eqn:rsq})) as a function of the location of partner fish relative to the direction of motion and position of a focal fish type (front fish or back fish). Focal fish are located at the origin of each plot, moving in the direction of the positive $x$-axis. Arrows in the top two panels indicate the mean direction of motion of partner fish. (C) contains the difference obtained from the front fish focused map minus the back fish focused map (for $R$ only).}\label{fig:relorientation_fxy}
\end{figure}

\begin{figure}[!h]
	\begin{center}
	\includegraphics[width=\textwidth]{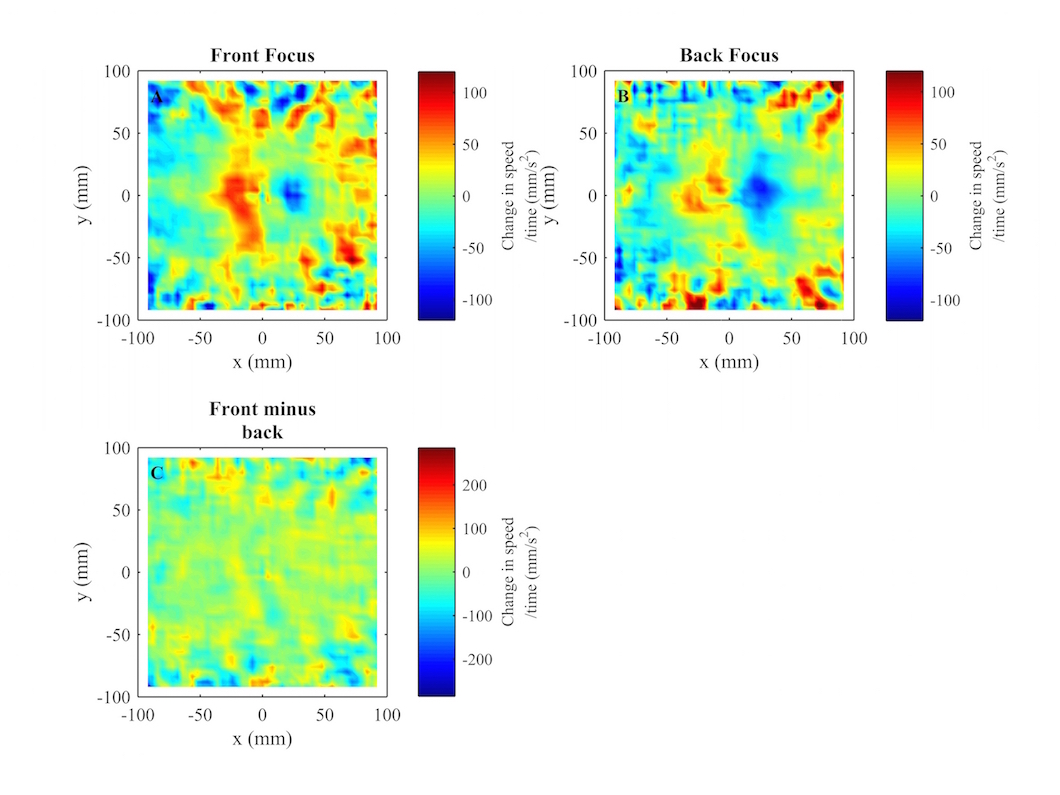}
	\end{center}
	\caption{(A) and (B) show heat maps of mean change in speed $(\textrm{mm}/s^{2})$ (the instantaneous change in magnitude of a fish's velocity vector) as a function of the location of their neighbour relative to the direction of motion and position of a focal fish type (front fish or back fish). Focal fish are located at the origin of each plot, moving in the direction of the positive $x$-axis. Extreme values for the mean change in speed have been truncated at $\pm 120 \textrm{mm}/\textrm{s}^{2}$ in the top two panels to allow for better visualisation of the details of lower magnitude changes in speed. (C) contains the difference obtained from the front fish focused map minus the back fish focused map.}\label{fig:changeinspeed_fxy}
\end{figure}

\begin{figure}[!h]
	\begin{center}
	\includegraphics[width=\textwidth]{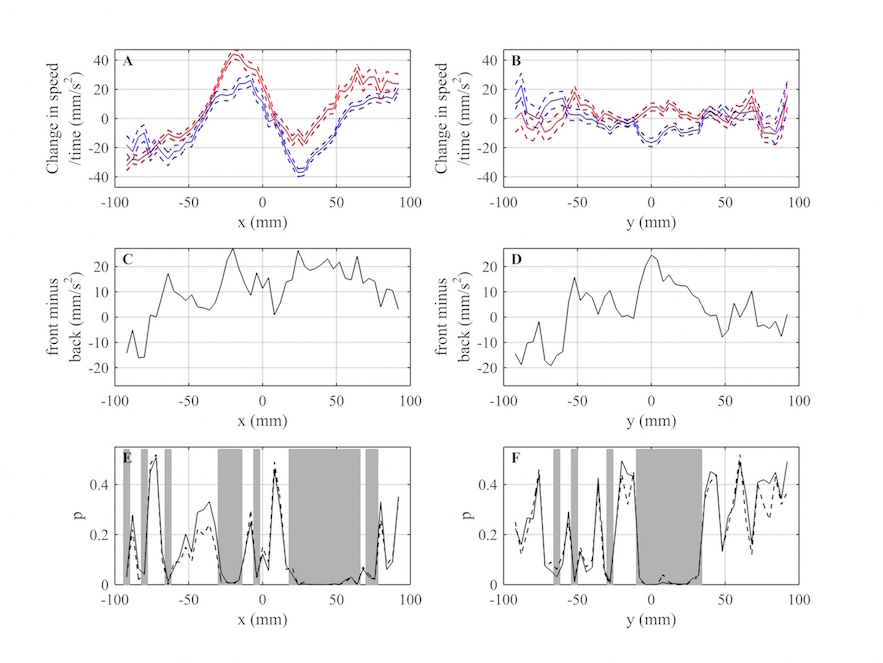}
	\end{center}
	\caption{Mean change in speed over time of focal fish as a function of their partner's $x$- (A) or $y$-coordinate (B). Curves representing the mean change in speed of front fish are plotted in red; curves representing the mean change in speed of back fish are plotted in blue. Dashed lines are plotted one standard error above and below the mean curves. (C) and (D) illustrate the difference in front fish and back fish curves, and (E) and (F) illustrate the estimated probability of a difference at least the same magnitude and sign as that observed occurring. Probabilities estimated from a set of 100 randomisation processes are plotted as dashed lines and probabilities derived from 1000 randomisations are plotted as solid lines. Grey regions indicate where estimated probabilities derived from 1000 randomisation processes are less than 0.05. Front fish and back fish differ in how they adjust their speed when their neighbour is close to them.}\label{fig:changeinspeed_projections}
\end{figure}

\begin{figure}[!h]
	\begin{center}
	\includegraphics[width=\textwidth]{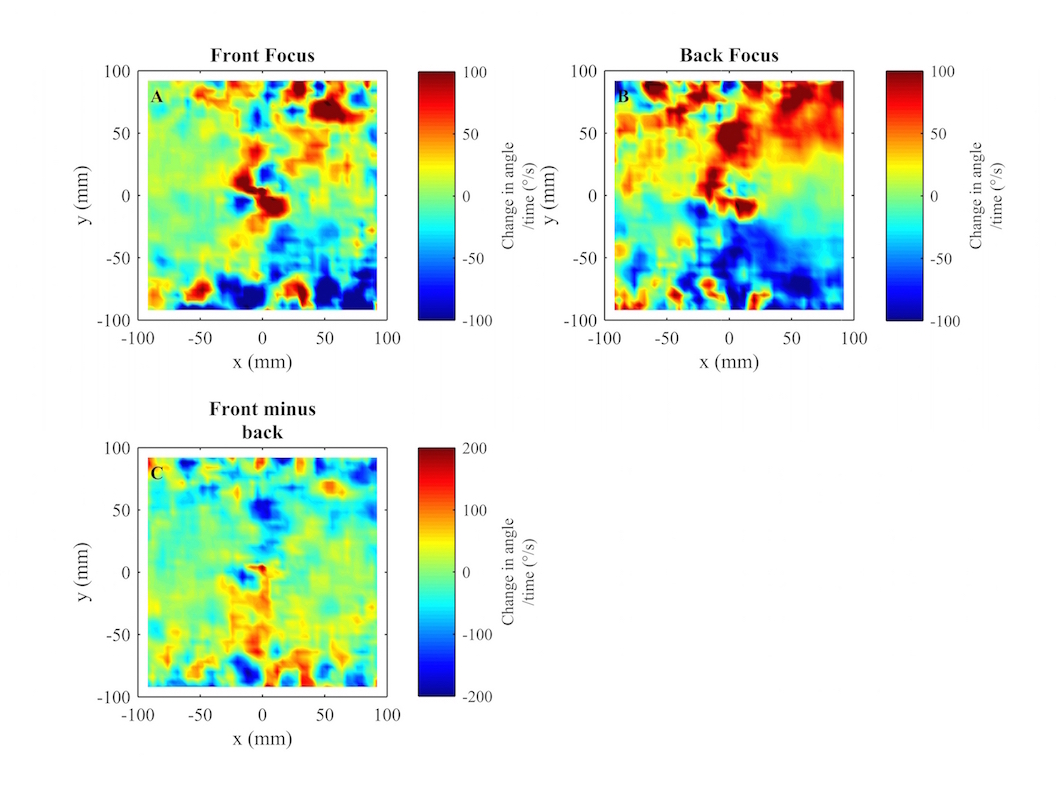}
	\end{center}
	\caption{(A) and (B) show heat maps of change in angle of motion over time (degrees/s) (the instantaneous change in direction of a fish's velocity vector) as a function of the location of partner fish relative to the direction of motion and position of a focal fish type (front fish or back fish). Focal fish are located at the origin of each plot, moving in the direction of the positive $x$-axis. Extreme values for the mean turning speed have been truncated at $\pm 100$ degrees/s in the top two panels to allow for better visualisation of the details of lower magnitude turning speeds. (C) shows the difference obtained from the front fish focused map minus the back fish focused map. Back fish have higher turning speeds towards their partner than front fish.}\label{fig:changeinangle_fxy}
\end{figure}

\begin{figure}[!h]
	\begin{center}
	\includegraphics[width=\textwidth]{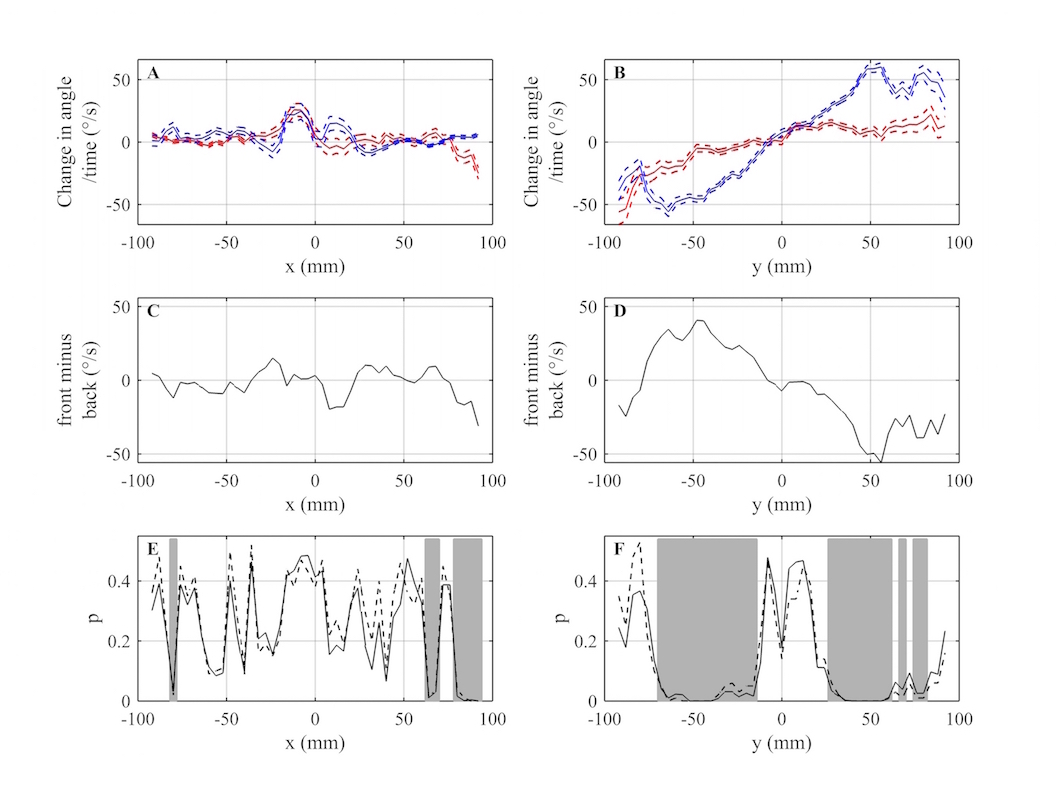}
	\end{center}
	\caption{Mean change in angle of motion over time of focal fish as a function of their partner's $x$- (A) or $y$-coordinate (B). Curves representing the mean turning speed of front fish are plotted in red; curves representing the mean turning speed of back fish are plotted in blue. Dashed lines are plotted one standard error above and below the mean curves. Positive values of turning speed indicate are associated with anti-clockwise turns; negative values are associated with clockwise turns. (C) and (D) illustrate the difference in front fish and back fish curves, and (E) and (F) illustrate the estimated probability of a difference at least the same magnitude and sign as that observed occurring. Probabilities estimated from a set of 100 randomisation processes are plotted as dashed lines and probabilities derived from 1000 randomisations are plotted as solid lines. Grey regions indicate where estimated probabilities derived from 1000 randomisation processes are less than 0.05.}\label{fig:changeinangle_projections}
\end{figure}

\begin{figure}[!h]
	\begin{center}
	\includegraphics[width=\textwidth]{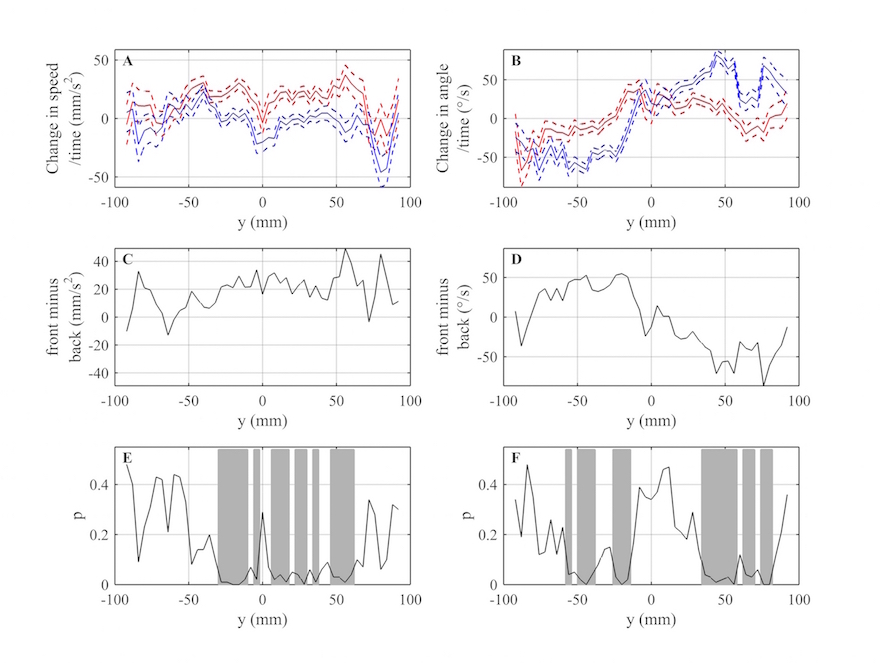}
	\end{center}
	\caption{The mean change in speed over time (A) and mean change in angle of motion over time (B) of front fish (red) and back fish (blue) projected onto the $y$-axis for the range of $x$-coordinates where fish were approximately next to each other ($-32 < x \leq 32$ mm). Dashed lines are plotted one standard error above and below the mean curves. (C) and (D) illustrate the difference in front fish and back fish curves, and (E) and (F) illustrate the estimated probability of a difference at least the same magnitude and sign as that observed occurring. Probabilities were estimated from a set of 100 randomisation processes here; grey regions indicate where these estimated probabilities are less than 0.05.}\label{fig:xmid40}
\end{figure}

Figures \ref{fig:propencounters_projections} to \ref{fig:changeinangle_projections} illustrate the proportion of encounters, speed, relative direction of motion of partner fish, change in speed over time and change in angle of motion over time for fish that occupied the front or back of their pair for the longest duration as a function of relative neighbour location. Probabilities determined to examine the likelihood that the sign and magnitude of observed differences between front fish and back fish associated quantities might appear through random assignation of front fish and back fish tags were similar irrespective of being derived from either the set of 100 random allocations or the set of 1000 random allocations. Figure \ref{fig:xmid40} illustrates the change in speed over time and the change in angle of motion over time of front and back fish projected onto the $y$-axis when partner fish are approximately beside a given focal fish (such that $-32 < x \leq 32$ (mm), a maximum difference between the approximate centres of the fish of approximately one body length).    

\end{document}